\documentclass[]{jfm}

\usepackage{graphicx}
\usepackage{newtxtext} 
\usepackage{newtxmath}
\usepackage{natbib}
\usepackage{tabularx}
\usepackage{multirow}
\usepackage{hyperref}
\usepackage{xcolor}
\usepackage{soul}
\usepackage{ulem}
\usepackage[final]{changes}

\hypersetup{
    colorlinks = true,
    urlcolor   = blue,
    citecolor  = blue,
}

\renewcommand{\Rey}{\mathit{Re}}
\newcommand{\Fr}{\mathit{Fr}}

\renewcommand{\vec}[1]{\boldsymbol{#1}}

\newcommand{\pder}[3]{\frac{\partial^{#1} {#2}}{\partial {#3}}}

\newcommand{\av}[1]{\langle #1\rangle}

\newcommand{\RomanNumeralCaps}[1]

\begin{document}

\title{Sidewall effects on the onset of interfacial Holmboe waves in stratified exchange flows at high Schmidt number}

\author{Guilherme Siqueira de Aquino\aff{1}, Metten M. de Lange\aff{2}, Adrien Lefauve\aff{3,4}
  \and Matias Duran-Matute\aff{2}\corresp{email address for correspondence: m.duran.matute@tue.nl}}

\affiliation{\aff{1}Univ Toulouse, Toulouse INP, CNRS, IMFT, Toulouse, France\\
\aff{2}Department of Applied Physics and Science Education, Eindhoven University of Technology, The Netherlands\\
\aff{3}Grantham Institute -- Climate Change and the Environment, Imperial College London, UK\\
\aff{4}Department of Civil and Environmental Engineering, Imperial College London, UK}
\maketitle

\abstract{Predicting the onset of interfacial instabilities is central to understanding turbulent mixing in natural and engineered stratified shear flows. Here, we study the onset of travelling Holmboe waves in confined exchange flows along a slope. Earlier stratified inclined duct experiments have mapped this transition, and stability analyses based on measured or prescribed profiles have explained their emergence. However, a predictive criterion linking forcing, geometry, base flow, and wave onset was still lacking. We closed this gap with a long-duct, sharp-interface asymptotic theory for the three-dimensional laminar exchange flow, including the effects of sidewall friction. The resulting analytical solution naturally identifies a confinement-adjusted Froude number, $Fr^*$, which unifies the effects of forcing and confinement into a single measure of the effective laminar exchange flow. Using this solution to parameterise sidewall drag in a practical width-averaged model, we perform linear stability analyses and numerical simulations at high Schmidt number. Together, linear stability analysis, direct numerical simulations, and existing experiments across a range of duct widths show that wave onset is accurately predicted by an approximately constant critical value of $Fr^*$. Deviations arise only in very narrow ducts, where sidewalls influence instability not only by modifying the laminar exchange flow but also by directly damping perturbations and delaying wave onset. These findings provide a predictive criterion for wave onset, reconcile long-standing discrepancies among experimental configurations, and establish lateral confinement as a fundamental control on the transition from laminar exchange to wave-driven mixing in stratified shear flows.}


\section{Introduction} \label{sec:intro}

Buoyancy-driven exchange flows occur whenever fluids of different density pass through a confined connection and are encountered in environments as diverse as estuarine salt wedges \citep{Geyer2010} and fluid mud \citep{Tu2022}, fjords \citep{Meire2023}, straits \citep{Farmer1988,Gregg1999}, ship locks \citep{Weiler2026,raaghav2025bubble}, and nuclear reactor cooling systems \citep{Leach1975,Mercer1975}. Many of these flows consist of two counter-flowing layers separated by a sharp density interface. Determining the conditions under which this base state becomes unstable is essential, because the resulting interfacial waves transport momentum and often constitute the first route to turbulence and irreversible mixing via wave breaking and subsequent secondary instabilities \citep{Peltier2003,Caulfield2021,Lefauve2025}. Predicting the onset of instability remains challenging because it depends not only on the imposed density contrast and forcing conditions, but also on the geometry of the connecting passage, particularly its length and degree of lateral confinement. 

\begin{figure}
    \centering    
    \includegraphics[width=0.6\linewidth]{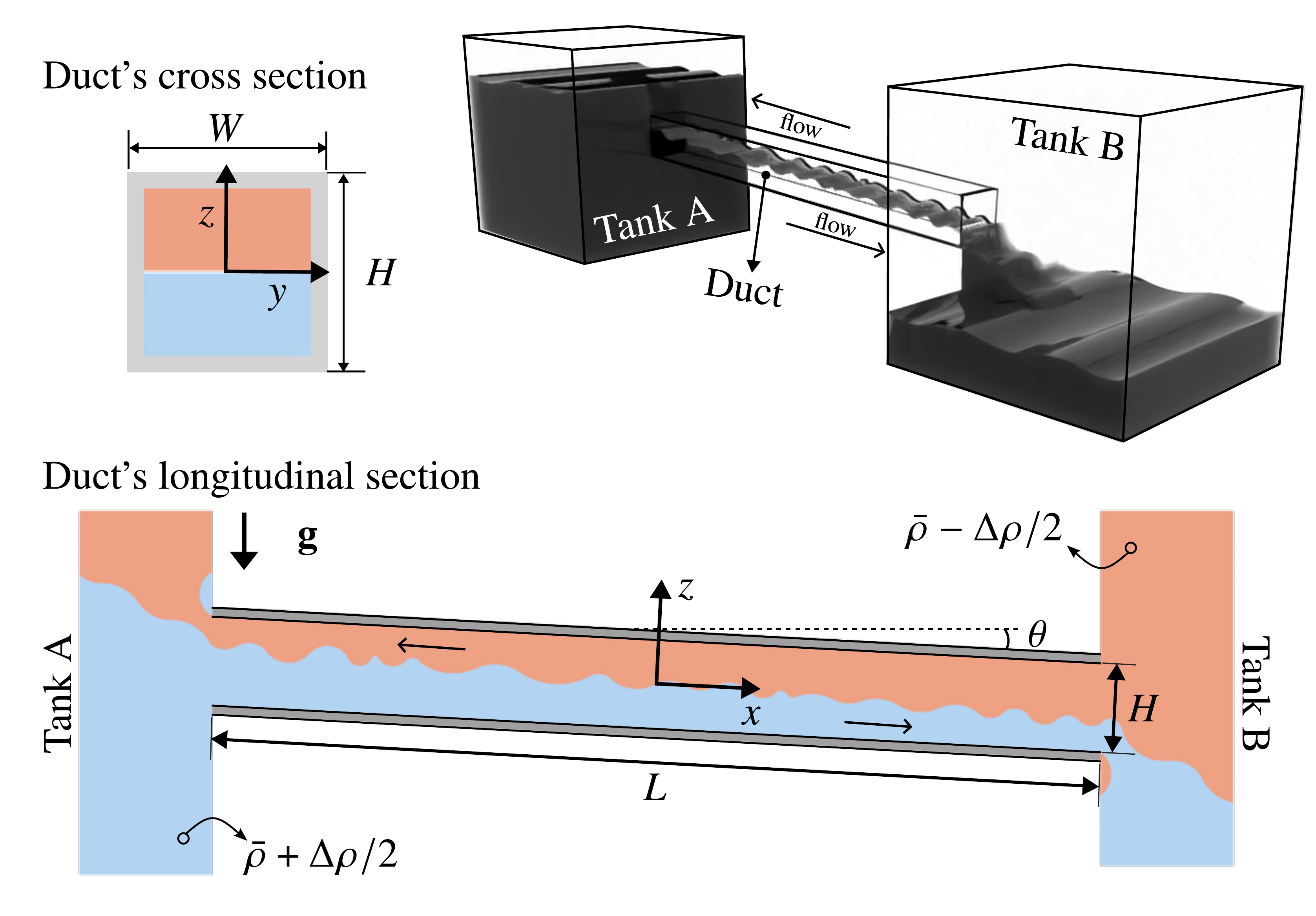}
    \caption{Schematic of the stratified inclined duct (SID) setup. Two reservoirs, Tank A and Tank B, are connected by a rectangular duct of height $H$, width $W$, and length $L$, inclined at an angle $\theta$. Initially, Tank A contains the denser fluid with density $\bar{\rho} + \Delta\rho/2$, while Tank B contains the lighter fluid with density $\bar{\rho} - \Delta\rho/2$. The denser fluid flows from Tank A to Tank B in the lower layer, and the lighter fluid flows from Tank B to Tank A in the upper layer. The coordinate system is centred at the duct midpoint, with $x$ along the duct axis, $y$ in the spanwise direction, and $z$ across the duct height.}
    \label{fig:setup}
\end{figure}

The stratified inclined duct (SID) is a canonical laboratory configuration to isolate these effects \citep[for a review, see][]{Lefauve2024review}. The SID consists of a long rectangular duct connecting two reservoirs and inclined at a small angle to the horizontal (figure~\ref{fig:setup}). The density difference between reservoirs is usually achieved by using water with different salinity or temperature. The resulting exchange flow is continuously forced by gravity and can be sustained for long times, allowing a controlled study of stratified shear flows from laminar motion to interfacial waves, intermittent turbulence, and sustained turbulence. Past experimental campaigns mapped these regimes in terms of the imposed tilt angle, duct geometry, and buoyancy Reynolds number (to be defined in the next section) \citep{Meyer2014,Lefauve2019,Lefauve2020}. Subsequent experimental studies characterised their mean flows and self-organisation \citep{Lefauve2022_exp1}, energetics and anisotropy \citep{Lefauve2022_exp2}, coherent vortical structures \citep{Lefauve2018,Jiang2022}, and temporal intermittency \citep{Lefauve2025}, providing a progressively richer picture of continuously forced, shear-driven stratified turbulence.

A long-standing objective of SID research has been to develop predictive criteria for regime transitions, dating back at least to \citet{Macagno1961}. In addition to being a reflection of our understanding of the flow dynamics, such criteria would help design future experimental campaigns targeting laminar, wave, or turbulent states without the need for extensive preliminary regime scans. \citet{Lefauve2019} showed that, at sufficiently large inclination, the transitions are organised by the product of the tilt angle and the Reynolds number, reflecting the balance between gravitational power input and dissipation. \citet{Duran-Matute2023} later linked horizontal and inclined ducts by deriving an analytical laminar solution in a long-duct, sharp-interface approximation. In physical terms, this approximation assumes that along-duct momentum is governed mainly by hydrostatic pressure, buoyancy forcing, and viscous drag, and that density is transported along the interface rather than across it by diffusion. It was further shown that the transitions are organised in the parameter space along isolines of the dimensionless volume flux, i.e. the Froude number, of the analytical solution.

However, the main limitation of the flux-based criterion proposed by \citet{Duran-Matute2023} is that it is strictly two-dimensional (2D): it does not account for variations of the exchange flow across the duct width, nor for the way sidewall friction modifies the velocity field, deforms the density interface, and affects the volume flux. These omissions matter because many geophysical and industrial exchange flows are confined laterally, and experiments with different duct aspect ratios show different transition thresholds \citep{Lefauve2020,Duran-Matute2023}. Hence, the flux-based criterion explains much of the transition behaviour, but not why the onset of waves differs across experimental configurations. This missing piece suggests that lateral confinement is not merely a secondary correction to the exchange flow, but a crucial ingredient in determining flow stability.

For large Schmidt number values, where the scalar diffusion is weak compared to the momentum diffusion, the interfaces remain sharp and the interfacial waves in SID are Holmboe waves \citep{Lefauve2018}. Notably, \citet{Lefauve2020} observed Holmboe waves in salt-stratified exchange flows but not in otherwise comparable temperature-stratified cases operating at a Schmidt number value two orders of magnitude lower. Since the seminal work of \cite{Holmboe1962OnLayers}, Holmboe waves have been the subject of extensive experimental, numerical, and theoretical investigations \citep{Hazel1972,Carpenter2007,Carpenter2010,Tedford2009}, and have also been reported in sediment-stratified estuarine lutoclines based on hydroacoustic observations \citep{Held2019}. However, theoretical studies on their onset and the stability of stratified shear flows with thin interfaces typically rely on simplifications that are not applicable in the SID. For example, most early studies have considered one-dimensional or inviscid flows \citep{Holmboe1962OnLayers,Hazel1972,SmythPeltier1989,Lawrence1991,ortiz2002spatial,haigh1999symmetric,Carpenter2011}, while the flow in the SID varies in all directions and is highly influenced by viscous forces. More recently, \citet{Lefauve2018} showed that the phase speeds, growth rates, and modal structure of the Holmboe waves are well predicted by a 3D linear stability analysis of the measured mean flow. \citet{Ducimetiere2021EffectsInstabilities} then isolated the effect of lateral confinement, showing that lateral walls generally reduce growth rates, modify mode selection, and introduce a family of spanwise-dependent modes. These studies established the importance of confinement, but relied on prescribed or measured base flows rather than deriving them from first principles and the imposed SID parameters.

Recent simulations and hydraulic analyses have clarified complementary SID transition mechanisms. \citet{Zhu2023} reproduced the experimentally observed laminar, wave, intermittent, and turbulent regimes in 3D direct numerical simulations (DNS). However, due to computational limitations, these were in the more diffusive temperature-stratified situation. \citet{Atoufi2023SID} used these simulation data to interpret SID transitions through two-layer hydraulics, showing how increasing tilt can drive the flow from subcritical and stable to supercritical and unstable to long waves, internal jumps, travelling waves, wave breaking, and intermittent turbulence. \citet{Zhu2024LongWave} also identified `long-wave' instabilities of sloping stratified exchange flows that are distinct from classical Kelvin–Helmholtz and Holmboe modes, which fall into the `short wave' category (i.e. not much longer than the duct height). These studies reveal distinct hydraulic and long-wave routes to turbulence. The present study addresses a different but complementary question: the onset of `short' Holmboe waves in large-Schmidt-number, laterally confined exchange flows with sharp density interfaces.

What remains missing is a predictive framework that connects the imposed forcing and duct geometry to the confined laminar base flow and subsequently to the onset of Holmboe instability. We provide such a framework by extending the long-duct, sharp-interface theory by \citet{Duran-Matute2023} to a 3D semi-analytical base flow including sidewall effects. This solution predicts the velocity field, density interface, and volume flux from the imposed control parameters, including the width-to-height ratio, and naturally identifies a confinement-adjusted Froude number that combines the effects of forcing and lateral confinement into a single measure of the laminar exchange flow. We then use a width-averaged formulation, parameterising the leading effect of sidewall friction, to perform linear stability analyses and large-Schmidt-number two-dimensional simulations. This framework allows us to determine whether the confinement-adjusted Froude number provides a predictive criterion for the onset of Holmboe waves, to quantify the distinct roles of sidewalls in the laminar base flow and the instability itself, and to reconcile the transition thresholds reported across previous SID experiments.

The remainder of the paper is organised as follows. Section~\ref{sec:problem} introduces the SID setup and governing parameters. Section~\ref{sec:3Dsolution} derives the 3D semi-analytical base-flow solution and quantifies the influence of sidewalls on the laminar exchange flow. Section~\ref{sec:wallEffects} introduces the width-averaged formulation and uses it for linear stability analysis. Section~\ref{sec:numericalResults} presents fully nonlinear simulations and compares the predicted wave onset with previous laboratory observations. Section~\ref{sec:conclusion} summarises the main findings.

\section{Problem description and governing parameters}\label{sec:problem}

The SID, shown schematically in figure~\ref{fig:setup}, consists of a rectangular duct inclined at an angle $\theta$ that connects two tanks filled with fluids of different density. We denote by $\bar{\rho}$ the mean density of the fluid in the system and by $\Delta \rho$ the initial density difference between the two tanks, so that the tank densities are $\bar{\rho}\pm\Delta\rho/2$. In this configuration, the denser fluid flows from Tank A to Tank B and the lighter fluid flows in the opposite direction, forming a stratified exchange flow. The tanks are large enough to be considered effectively infinite over the duration of the experiments and numerical simulations. The duct has length $L$, height $H$, and width $W$. The coordinate system is chosen with $x$ along the duct, $y$ in the spanwise direction, and $z$ across the duct height, as shown in figure~\ref{fig:setup}.

Five dimensionless numbers govern the flow. The first three are geometrical: the tilt angle of the duct $\theta$, and the longitudinal and spanwise aspect ratios
\begin{equation}
A\equiv \frac{L}{H}\quad \text{and}\quad B\equiv \frac{W}{H},
\end{equation}
respectively. The aspect ratio $A$ is related to the small internal angle $\alpha$ shown in figure~\ref{fig:setup} by $A=\cot\alpha$, which reduces to $A\approx \alpha^{-1}$ for long ducts with $A\gg1$. We refer to $B$ as the confinement parameter: $B\ll1$ corresponds to narrow ducts, whereas $B\gg1$ corresponds to wide ducts. The fourth and fifth parameters are the buoyancy or gravitational Reynolds number and the Schmidt number,
\begin{equation}
\Rey_g\equiv\frac{HU_g}{2\nu}\quad \text{and}\quad  Sc\equiv \frac{\nu}{\kappa},
\label{eq:Reynolds}
\end{equation}
where $\nu$ is the kinematic viscosity, $U_g\equiv\sqrt{g'H}$ is the buoyancy velocity scale, and $g'=g\Delta\rho/\bar{\rho}$ is the reduced gravity with $g$ the gravitational acceleration. The buoyancy Reynolds number characterises the relative strength of buoyancy-driven inertial forces compared with viscous forces. The Schmidt number compares molecular viscous momentum diffusion with the molecular diffusivity $\kappa$ of the scalar responsible for the density difference. For heat, the analogous parameter is more commonly called the Prandtl number. The two most common values in SID experiments are $Sc\approx700$ for salt in water and $Sc\approx7$ for heat in water \citep{Lefauve2020}. In this work, we consider only large $Sc$ values relevant to salt-stratified water and other slowly diffusing constituents.

To quantify the flow response, we use the Froude number
\begin{equation}
\Fr\equiv\frac{U}{U_g}
=\frac{1}{U_g HW}\int^{H/2}_{-H/2}\int^{W/2}_{-W/2}\left| u\right|\,\mathrm{d}y\, \mathrm{d}z,
\label{eq:Froude3D}
\end{equation}
where $u$ is the streamwise velocity and $U$ is its cross-sectional mean magnitude. This parameter is the dimensionless volume flux through the duct and is therefore also often denoted by $Q$, for example by \citet{Lefauve2019}.

The problem considered here is to determine the ranges of the control parameters $(\Rey_g,\theta,B)$ for which the flow becomes unstable and supports Holmboe waves, under the assumptions of long ducts, $A\gg1$, and weak scalar diffusion, $Sc\gg1$. Within this setting, we also quantify how the occurrence of Holmboe waves affects the volumetric flow rate through the duct, as measured by $\Fr$.

\section{Three-dimensional solution for the base laminar flow} \label{sec:3Dsolution}

This section derives the laminar base flow used throughout the rest of the paper. For each imposed geometry and forcing, the goal is to determine three quantities: the 3D streamwise velocity field $u$, the interface shape along the streamwise direction $\eta$, and the corresponding laminar volume flux $\Fr^*$. Hereafter, $\Fr^*$ denotes the theoretical prediction of the Froude number for the laminar flow, while $\Fr$ denotes the actual value obtained in experiments or numerical simulations. The new element relative to the 2D theory of \citet{Duran-Matute2023} is the explicit dependence on the lateral confinement $B$. This dependence enters through sidewall friction and modifies both the base flow and the interface shape that serve as the base state for the stability analysis in later sections.

\subsection{Derivation of the 3D semi-analytical solution}

We seek a steady, laminar, sharp-interface solution in the long-duct limit $A\gg1$ and at high Schmidt number $Sc\gg 1$. In this limit, the leading-order balance is hydrostatic in the vertical, viscous in the momentum equations, and advective in the density equation. The derivation below identifies the resulting forced Poisson problem for the streamwise velocity and shows how sidewall friction enters through the confinement parameter $B$.

We consider the flow to be governed by the Navier--Stokes equations, the continuity equation for an incompressible fluid, and the advection--diffusion equation:
\begin{subequations}\label{eq:NS-before}
\begin{align}
    \pder{}{\vec{u}}{t}
    +\vec{u}\cdot \vec{\nabla}\vec{u}
    &=
    -\frac{1}{\bar{\rho}}\vec{\nabla}p
    +\nu \nabla^2 \vec{u}
    +\frac{\rho'}{\bar{\rho}}\vec{g},
    \label{eq:NS} \\
    \vec{\nabla}\cdot \vec{u}
    &=0,
    \label{eq:Continuity} \\
    \pder{}{\rho'}{t}
    +\vec{u}\cdot\vec{\nabla}\rho'
    &=
    \kappa\nabla^2 \rho',
    \label{eq:AdvectionDiffusion}
\end{align}
\end{subequations}
where $\vec{u}=(u, v, w)$ is the flow velocity with its components in the $x$-, $y$- and $z$-directions, respectively, $p$ is the pressure, $\rho'=\rho-\bar{\rho}$ is the deviation of the density $\rho$ from the mean and $\vec{g}=g(\sin\theta,0,-\cos\theta)$. Since we consider small values of the density difference with respect to the mean density, we use the Boussinesq approximation. 

To find the analytical solution using asymptotic analysis, we make \eqref{eq:NS-before} dimensionless for a steady flow. For this, we define the dimensionless tilde variables
\begin{equation}
\label{eq:scalings}
\begin{gathered}
    u=U\tilde{u},\ \
    v=\frac{B}{A}U\tilde{v},\ \
    w=\frac{1}{A}U\tilde{w},\ \
    x=\frac{L}{2}\tilde{x},\ \
    y=\frac{W}{2}\tilde{y},\ \
    z=\frac{H}{2}\tilde{z},\ \
    \rho'=\frac{\Delta\rho}{2}\tilde{\rho}',\ \
    p=\frac{2\bar{\rho}U\nu L}{H^2}\tilde{p}.
\end{gathered}
\end{equation}
Substituting the scalings \eqref{eq:scalings} into \eqref{eq:NS-before} gives the steady dimensionless equations reported in Appendix~\ref{sec:AppendixSolution}. The leading-order balance in the long-duct limit is the only part needed here. We now consider the long-duct limit, $A\to\infty$, while all other parameters remain finite. This allows us to perform an asymptotic analysis using $A^{-1}$ as the small parameter.

The velocities, the variable part of the density, and the pressure are written as asymptotic expansions according to 
\begin{equation}
    \tilde{u}=\sum^\infty_{n=0} A^{-n}\tilde{u}_n, \quad
    \tilde{v}=\sum^\infty_{n=0}A^{-n}\tilde{v}_n, \quad
    \tilde{w}=\sum^\infty_{n=0} A^{-n}\tilde{w}_n, \quad
    \tilde{p}=\sum^\infty_{n=0} A^{-n}\tilde{p}_n,  \quad
    \tilde{\rho}'=\sum^\infty_{n=0} A^{-n}\tilde{\rho}'_n.
    \label{eq:AsymptoticExpansions}
\end{equation}
Keeping only the leading-order terms gives the hydrostatic/gravitational/viscous momentum balance and the advective density balance (HGV-A) \citep{Duran-Matute2023}, also known as the high-advection/low-diffusion approximation for laminar flows in horizontal ducts \citep{Kaptein2020}. Streamwise forcing is balanced by viscous diffusion, the vertical momentum equation is hydrostatic, and scalar diffusion is negligible, such that the corresponding zeroth-order equations are
\begin{subequations}
\label{eq:HGV-A}
\begin{align}
    \frac{\partial \tilde{p}_0}{\partial \tilde{x}} 
    &=
    \underbrace{B^{-2} \frac{\partial^2 \tilde{u}_0}{\partial \tilde{y}^2}}_{\text{confinement effects}}
    + \frac{\partial^2 \tilde{u}_0}{\partial  \tilde{z}^2}
    + \frac{K}{4}A \tilde{\rho}_0'\sin{\theta},
    \label{eq:hydrostatic_x}\\[1.0ex]
    \frac{\partial \tilde{p}_0}{\partial \tilde{y}}
    &=0, \\[1.0ex]
    \frac{\partial \tilde{p}_0}{\partial \tilde{z}} 
    &= -\frac{K}{4}\tilde{\rho}_0'\cos{\theta},
    \label{eq:Hydrostatic_z} \\[1.0ex]
    \frac{\partial^2 \tilde{\rho}_0'}{\partial \tilde{z}^2} 
    &= 0,
    \label{eq:Transport_assym}
\end{align}
\end{subequations}
where $K$ is the dimensionless ``pressure--viscous'' parameter
\begin{equation}
    K\equiv \frac{\Rey_g}{A \, \Fr}=\dfrac{U_g^2 \, H^2}{2 \nu \, L \,  U},
    \label{eq:definitionK}
\end{equation}
which must remain finite in the long-duct limit. This parameter characterises the ratio of horizontal pressure-gradient forcing $U_g^2/L$ to vertical viscous diffusion $\nu U/H^2$, and is analogous to a Galilei-type number. The only new leading-order term relative to the 2D solution is $B^{-2}\partial^2\tilde{u}_0/\partial\tilde{y}^2$, which represents viscous momentum diffusion by the sidewalls.

We assume that the density is organised in two layers with a sharp interface between them,
\begin{equation}
    \tilde{\rho}_0'(\tilde{x}, \tilde{z})=1-2\mathcal{H}(\tilde{z}-\eta(\tilde{x})),
    \label{eq:density}
\end{equation}
where $\mathcal{H}$ is the Heaviside step function and $\tilde{z}=\eta(\tilde{x})$ is the interface location. We solve the leading-order equations separately in each layer, denoting the upper and lower layers by $\zeta=+1$ and $\zeta=-1$, respectively, so that $\tilde{\rho}'_{0,\zeta}=-\zeta$. The resulting velocity satisfies a forced Poisson problem in the $(\tilde{y},\tilde{z})$ plane, with no-slip conditions at the sidewalls, top and bottom walls, and interface (see Appendix \ref{app:poisson-solution} for more details).

The resulting streamwise velocity field is
\begin{equation}
\begin{split}
    \tilde{u}_{0,\zeta}(\tilde{x}, \tilde{y}, \tilde{z})=- \zeta K\frac{F_\zeta(\tilde{x})(1-\zeta \eta)^2}{\pi^3}\sum_{n=1}^\infty \frac{\sin{\left[\frac{\pi(2n-1) (\tilde{z}+a_\zeta)}{1-\zeta\eta}\right]}}{(2n-1)^3}
    \left\{\frac{\cosh\left[\frac{(2n-1)\pi B\tilde{y}}{1-\zeta\eta}\right]}{\cosh\left[\frac{(2n-1)\pi B}{1-\zeta\eta}\right]}-1\right\},
    \label{eq:EquationForuwithn}
\end{split}
\end{equation}
where $n\in\mathbb{N}^+$, and
\begin{equation}
a_\zeta = 
    \begin{cases}
        1 & \text{if } \zeta=-1,\\
        -\eta & \text{if } \zeta=+1.
    \end{cases}
\end{equation}
The forcing amplitude in each layer is
\begin{equation}
   \dfrac{K}{4} F_\zeta(\tilde{x})=-\frac{\pi ^4}{16} \frac{\left[(\zeta +1) (\eta +1)^3 P_{-1}(\tilde{x})+(\zeta -1) (\eta -1)^3 P_{+1}(\tilde{x})\right]}{\left(\eta ^2-1\right)^3 P_{-1}(\tilde{x})P_{+1}(\tilde{x}) },
    \label{eq:Fx}
\end{equation}
where
\begin{equation}
    P_\zeta(\tilde{x}) \approx -\frac{\pi^4}{96}+\frac{1-\zeta\eta}{B\pi}\tanh{\left(\frac{B\pi}{1-\zeta\eta}\right)}.
    \label{eq:Pzeta}
\end{equation}
Here, $P_\zeta$ has been approximated by retaining only the dominant $n=1$ term in the series given in Appendix~\ref{app:poisson-solution}. The expression above gives $\tilde{u}_0$ once the interface shape $\eta(\tilde{x})$ and the parameter $K$ are known. These are determined together from the autonomous equation
\begin{equation}
    \frac{\mathrm{d}\eta}{\mathrm{d}\tilde{x}}=-\frac{\pi^4}{4K}\frac{\left[-P_{+1}(\tilde{x})(-1+\eta)^3+P_{-1}(\tilde{x})(1+\eta)^3\right]}{P_{+1}(\tilde{x})P_{-1}(\tilde{x})\left(-1+\eta^2\right)^3\cos{\theta}}+A\tan{\theta}.
    \label{eq:detadx}
\end{equation}
This equation is solved with end conditions on $\eta$, as described next.

\subsection{Solving for the interface shape $\eta$ and determining $K$} \label{sec:interface}
We now determine the two remaining unknowns in the base-flow solution: the interface shape $\eta(\tilde{x})$ and the parameter $K$. Since SID experiments use small tilt angles, we take $\cos\theta\simeq1$ and $\sin\theta\simeq\theta$. Under this approximation, \eqref{eq:detadx} depends on the imposed geometry only through $A\sin\theta\simeq\theta/\alpha$ and $B$. Thus, for each pair $(A\sin\theta,B)$, solving for the interface reduces to finding the corresponding value of $K$.

\begin{figure}
    \centering
    \includegraphics[width=0.95\textwidth]{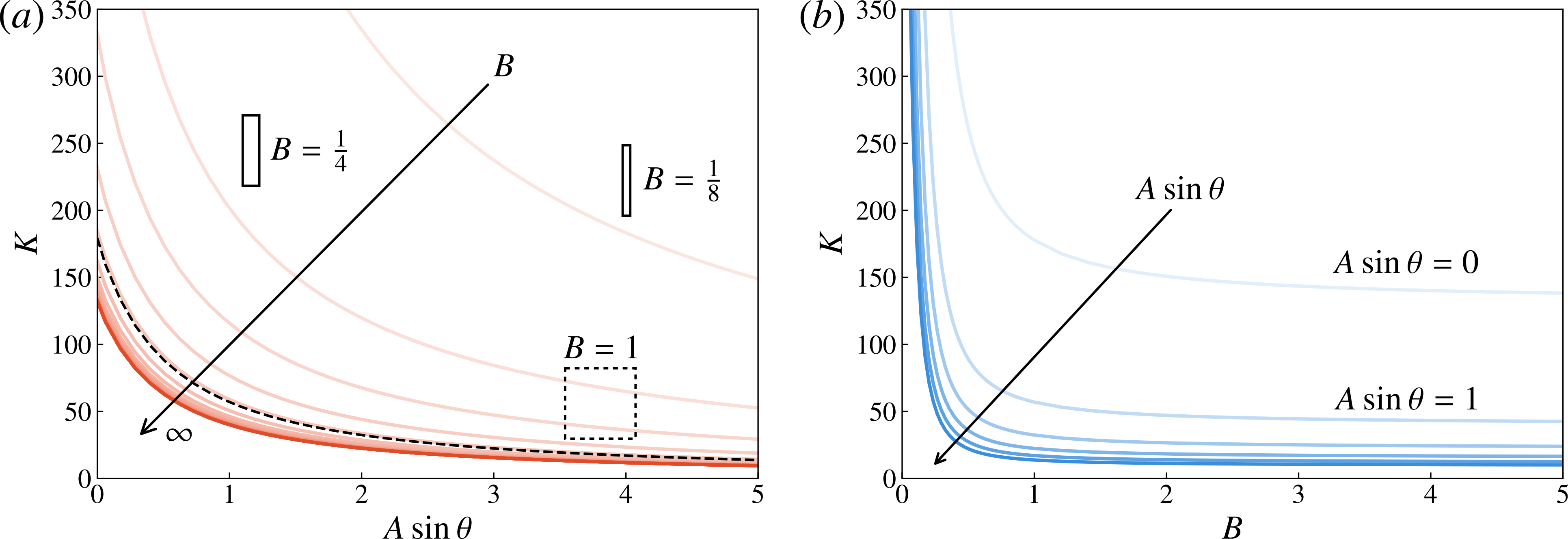}
     \caption{Dependence of the parameter $K$ on the imposed geometry in the 3D semi-analytical laminar solution. (\textit{a}) $K$ as a function of $A\sin\theta$ for spanwise aspect ratios $B\equiv W/H\in[1/8,1000]$. The dashed curve marks the case $B=1$ (i.e., a square duct). (\textit{b}) $K$ as a function of $B$ for fixed values of $A\sin\theta\in[0,5]$. Arrows indicate increasing $B$ in (\textit{a}) and increasing $A\sin\theta$ in (\textit{b}).}
    \label{fig:Kvalues}
\end{figure}

The value of $K$ is fixed by imposing end conditions on the interface. Following \citet{Kaptein2020}, we assume that the interface curves into thin entrance and exit currents near the duct ends with dimensionless thickness $\epsilon<A^{-1}\ll 1$. In the small-$\epsilon$ approximation, we simplify the end conditions to $\eta(\pm1)=\mp1$ to determine $\eta(\tilde{x})$ and $K$ for given values of $(A\sin\theta,B)$. Figure~\ref{fig:Kvalues} shows how the pressure--viscous forcing parameter $K$ depends on the imposed geometry. Narrower ducts have larger $K$, reflecting the increased importance of spanwise viscous diffusion. For $B\gtrsim1$, this sidewall contribution becomes weak, and the solution approaches the 2D limit of \citet{Duran-Matute2023}.

Figure~\ref{fig:Kvalues} also shows that $K$ decreases as $A\sin\theta$ increases. Physically, increasing $A\sin\theta$ strengthens the component of gravity along the duct relative to the pressure-gradient forcing; therefore, the pressure--viscous forcing parameter $K$ decreases. In the opposite geometric limit, $B\to0$, the solution gives $K\to\infty$, consistent with momentum diffusion being dominated by the sidewalls rather than by the top and bottom walls.

\begin{figure}
    \centering
    \includegraphics[width=0.95\textwidth]{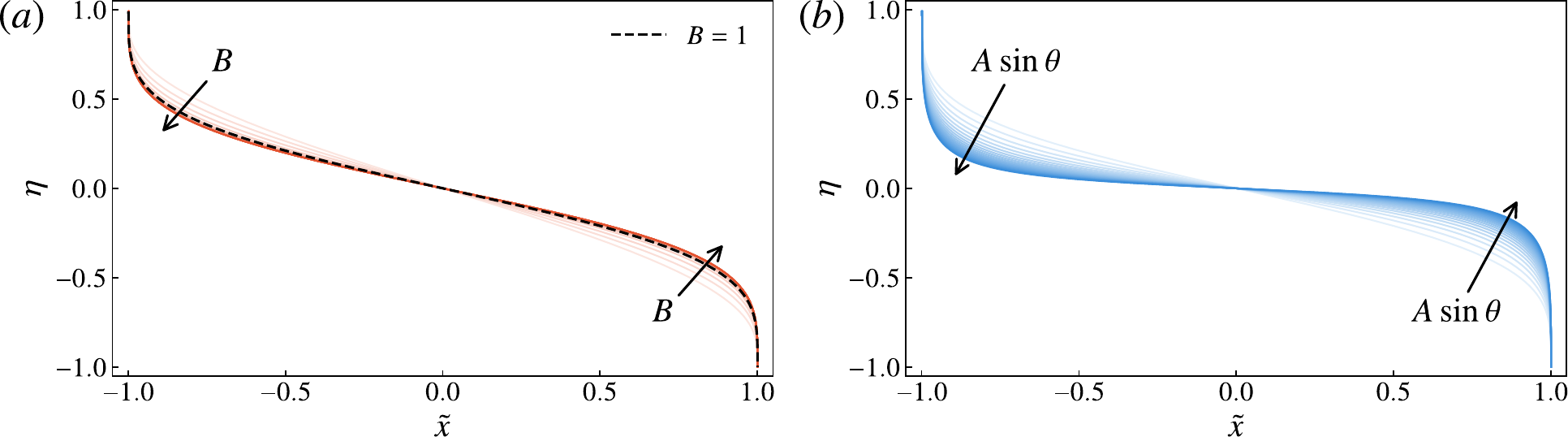}
    \caption{Dependence of the interface shape $\eta(\tilde{x})$ on confinement and relative inclination. (\textit{a}) Interface shape for fixed $A\sin\theta=1$ and varying spanwise aspect ratio $B\in[1/8,1000]$. (\textit{b}) Interface shape for fixed square duct ($B=1$) and varying $A\sin\theta\in[0,5]$. Arrows indicate increasing $B$ in (\textit{a}) and increasing $A\sin\theta$ in (\textit{b}), following the colour gradient from lighter to darker.}
    \label{fig:eta}
\end{figure}

Once $K$ is fixed, \eqref{eq:detadx} gives the interface shape $\eta(\tilde{x})$. Figure~\ref{fig:eta} shows how this shape varies with $B$ and $A\sin\theta$. For fixed $A\sin\theta$, the narrowing of the duct increases the slope of the interface near the duct centre, while wider ducts produce a flatter central interface. For fixed $B$, increasing $A\sin\theta$ also flattens the interface near the centre and concentrates the variation near the duct ends.

\subsection{Analytical velocity field: features and validation}

\begin{figure}
    \centering
    \includegraphics[width=0.7\textwidth]{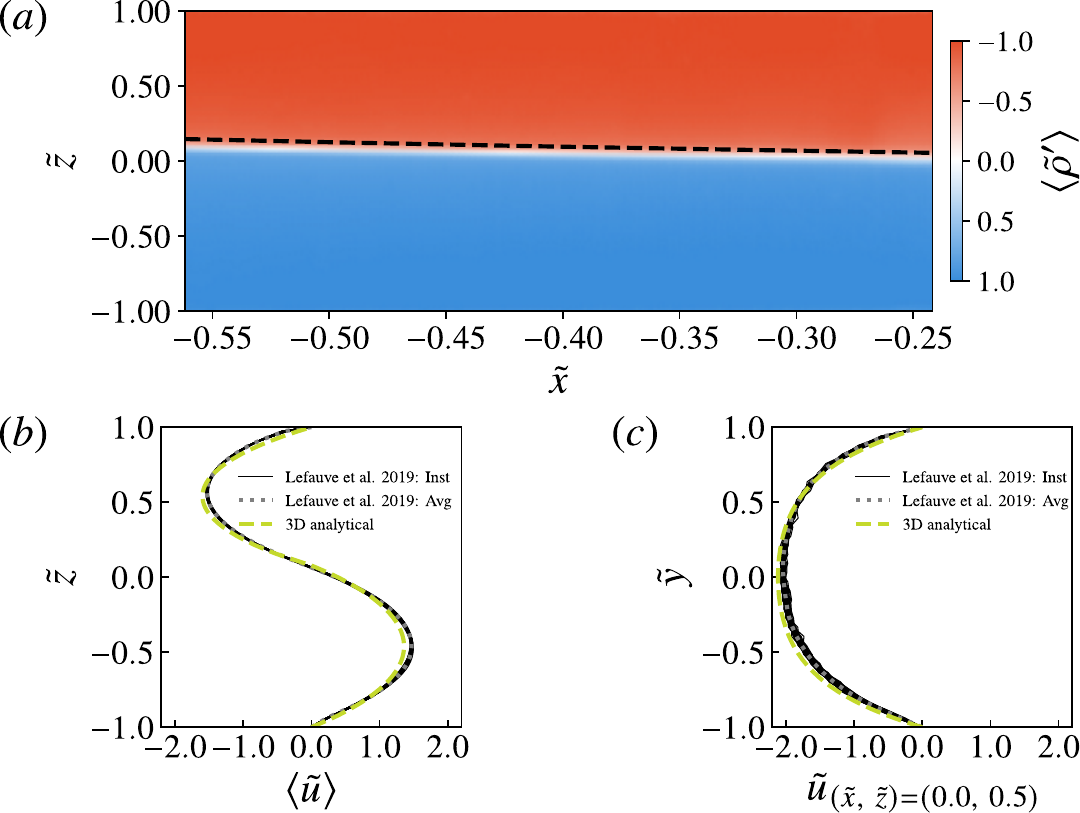}
    \caption{Agreement of laminar-flow characteristics at $(\Rey_g, A \sin \theta) = (398, 1.05)$ between experimental measurements by \citet{Lefauve2019} and the proposed 3D analytical model. (\textit{a}) Time- and width-averaged density field in the $(\tilde{x}, \tilde{z})$ plane; the dashed curve indicates the interface predicted by the semi-analytical solution. (\textit{b}) Experimental width-averaged profiles of the along-duct velocity compared with the width-averaged semi-analytical profile for $\tilde{x}=-0.33$. (\textit{c}) Experimental velocity profile along the spanwise direction compared with the semi-analytical profile at ($\tilde{x},\ \tilde{z}$) $=$ ($0.0,\ 0.5$). In (\textit{b}) and (\textit{c}), instantaneous experimental profiles are shown as solid black curves, and the corresponding time-averaged profile as a dotted curve.}
    \label{fig:AnalyticalValidation}
\end{figure}

We next validate the semi-analytical base flow against laminar SID measurements. The comparison uses the data of \citet{Lefauve2019}, made available by \citet{Lefauve2019b}, for a duct with $(A,B)=(30,1)$ and salt stratification at $Sc=700$. Figure~\ref{fig:AnalyticalValidation} confirms the sharp-interface and velocity predictions. In figure~\ref{fig:AnalyticalValidation}(\textit{a}), the predicted interface follows the measured contour where $\av{\tilde{\rho}'}\approx0$ (with $\av{\cdot}$ denoting the width average). This result supports the assumed two-layer density structure. Figure~\ref{fig:AnalyticalValidation} (\textit{b}) compares the predicted and measured width-averaged velocity profiles at $\tilde{x}=-0.33$. The model captures both the magnitude and the Poiseuille-like shape of the counter-flowing layers. The asymmetry of these profiles follows from the variation of $\eta(\tilde{x})$: as one layer thins toward an end of the duct, mass conservation increases its maximum velocity.

Figure~\ref{fig:AnalyticalValidation}(\textit{c}) provides an independent check of the spanwise structure by comparing $\tilde{u}_0$ at $(\tilde{x},\tilde{z})=(0.0,0.5)$. The measured and predicted profiles both have a flattened Poiseuille-like shape, confirming that the semi-analytical solution captures the 3D structure of the laminar base flow. This agreement supports its use as the base state for the stability analysis in the following section.

\begin{figure}
    \centering
     \includegraphics[width=0.8\linewidth]{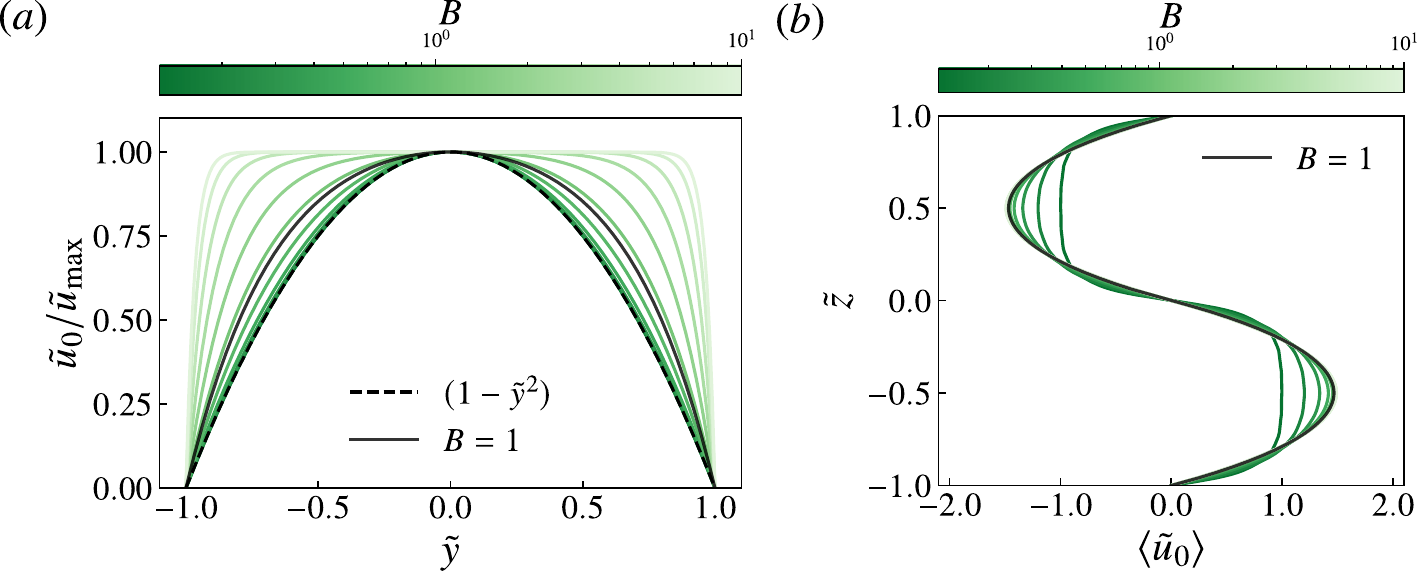}
    \caption{Effect of lateral confinement on the normalised streamwise velocity $\tilde{u}_0$. (\textit{a}) Spanwise profiles at $(\tilde{x},\tilde{z})=(0,0.5)$, normalised by the maximum velocity $\tilde{u}_{\max}$, for different values of $B$. The dashed curve shows the Poiseuille profile $\tilde{u}/\tilde{u}_{\max}=1-\tilde{y}^2$. (\textit{b}) Vertical profiles of $\av{ \tilde{u}_0}$ at $\tilde{x}=0$. The solid black curve in both panels marks the square duct case $B=1$.}\label{fig:profileyB}
\end{figure}

Figure~\ref{fig:profileyB} isolates the effect of lateral confinement on the normalised streamwise velocity $\tilde{u}_0$. Figure~\ref{fig:profileyB}(\textit{a}) shows the spanwise profiles at $(\tilde{x},\tilde{z})=(0,0.5)$, normalised by the local maximum $\tilde{u}_{\max}$, for varying $B$. For small $B$, the profile approaches the Poiseuille solution $\tilde{u}/\tilde{u}_{\max}=1-\tilde{y}^2$ (dashed line), showing that sidewall friction dominates the momentum balance. This influence remains substantial at $B=1$ (solid black line). For $B\gtrsim4$, the velocity becomes nearly uniform across the duct core, with sidewall effects confined to thin lateral boundary layers. Figure~\ref{fig:profileyB}(\textit{b}) shows the corresponding vertical profiles of $\langle\tilde{u}_0\rangle$ at the duct centre ($\tilde{x}=0$). As $B$ increases, these approach the expected two-layer Poiseuille-like structure.

The key result of the base-flow solution is the confinement-adjusted Froude number:
\begin{equation}
     \Fr^* \equiv \frac{\Rey_g}{AK},
     \label{eq:Fr*}
\end{equation}
which represents the dimensionless laminar volume flux and follows directly from the definition of $K$ in \eqref{eq:definitionK}. Unlike the 2D theory of \citet{Duran-Matute2023}, the present $\Fr^*$ includes the effect of the duct width through the $B$-dependence of $K$.

Equation~\eqref{eq:Fr*} shows that all imposed control parameters influence the magnitude of the laminar exchange flow through the single parameter $\Fr^*$. Increasing $\Rey_g$ increases $\Fr^*$, whereas increasing the duct length through $A$ decreases it. The remaining dependence on $B$ and $A\sin\theta$ enters through $K$ (recall figure~\ref{fig:Kvalues}): wider ducts and larger relative inclinations both reduce $K$, and therefore increase $\Fr^*$. In the limit $B\gg1$, the solution recovers the 2D approximation of \citet{Duran-Matute2023}. However, for finite $B$, the width dependence of $K$ provides the explicit confinement correction needed to compare ducts of different aspect ratios. Furthermore, since $\Fr^*$ only coincides with the actual value of $\Fr$ when the assumptions of the semi-analytical solution hold, deviations in $\Fr/\Fr^*$ indicate departures from the laminar sharp-interface base-flow model.

\section{Width-averaged framework and linear stability} \label{sec:wallEffects}

We now reduce the 3D confined base-flow problem to a width-averaged 2D problem. The aim is to retain the leading dynamical effect of the sidewalls while making the stability analysis and nonlinear simulations tractable across the parameter space $(\Rey_g,A \sin \theta, B)$ at high $Sc$. We do this in two steps. First, we use the 3D semi-analytical solution of \S\ref{sec:3Dsolution} to parameterise sidewall friction by an effective drag coefficient $C_D$. Second, we introduce this drag into the time-dependent width-averaged equations used for the linear stability analysis and simulations below.

\subsection{Parametrisation of the sidewall friction}

The first ingredient of the width-averaged model is an effective sidewall drag. We estimate this drag directly from the 3D semi-analytical laminar solution of \S\ref{sec:3Dsolution}, so that the model retains the leading effect of lateral confinement. We define the width-averaged base velocity in each layer as
\begin{equation}
    \av{u_{0,\zeta}} = \frac{1}{2}\int^1_{-1}\tilde{u}_{0,\zeta}\mathrm{d}\tilde{y}.
\end{equation}

Averaging the forced Poisson equation \eqref{eq:poisson} across the duct width gives
\begin{equation}
\label{eq:NS_width_av}
    \frac{\partial^2 \av{u_{0,\zeta}}}{\partial \tilde{z}^2} + B^{-2}\left.\frac{\partial \tilde{u}_{0,\zeta}}{\partial \tilde{y}}\right|_{\tilde{y}=1} = -\zeta \frac{K}{4} F_\zeta (\tilde{x}),
\end{equation}
where symmetry about $\tilde{y}=0$ has been used in the sidewall term. We approximate this term by a linear drag law, analogous to bottom-drag parameterisations in shallow flows \citep{Dolzhanskii_Krymov_Manin_1992,clercx2003quasi,duran2011scaling},
\begin{equation}
\label{eq:parametrisation_CD}
     B^{-2}\left.\frac{\partial \tilde{u}_{0,\zeta}}{\partial \tilde{y}}\right|_{\tilde{y}=1} \approx C_D \av{u_{0,\zeta}}
\end{equation}
where $C_D$ is an effective sidewall drag coefficient. The width-averaged base-flow equation then becomes
\begin{equation}
    \frac{\partial^2 \av{u_{0,\zeta}}}{\partial \tilde{z}^2} = -\zeta \frac{K}{4} F_\zeta (\tilde{x}) - C_D \av{u_{0,\zeta}}.
\end{equation}

\begin{figure}
    \centering
    \includegraphics[width=0.95\textwidth]{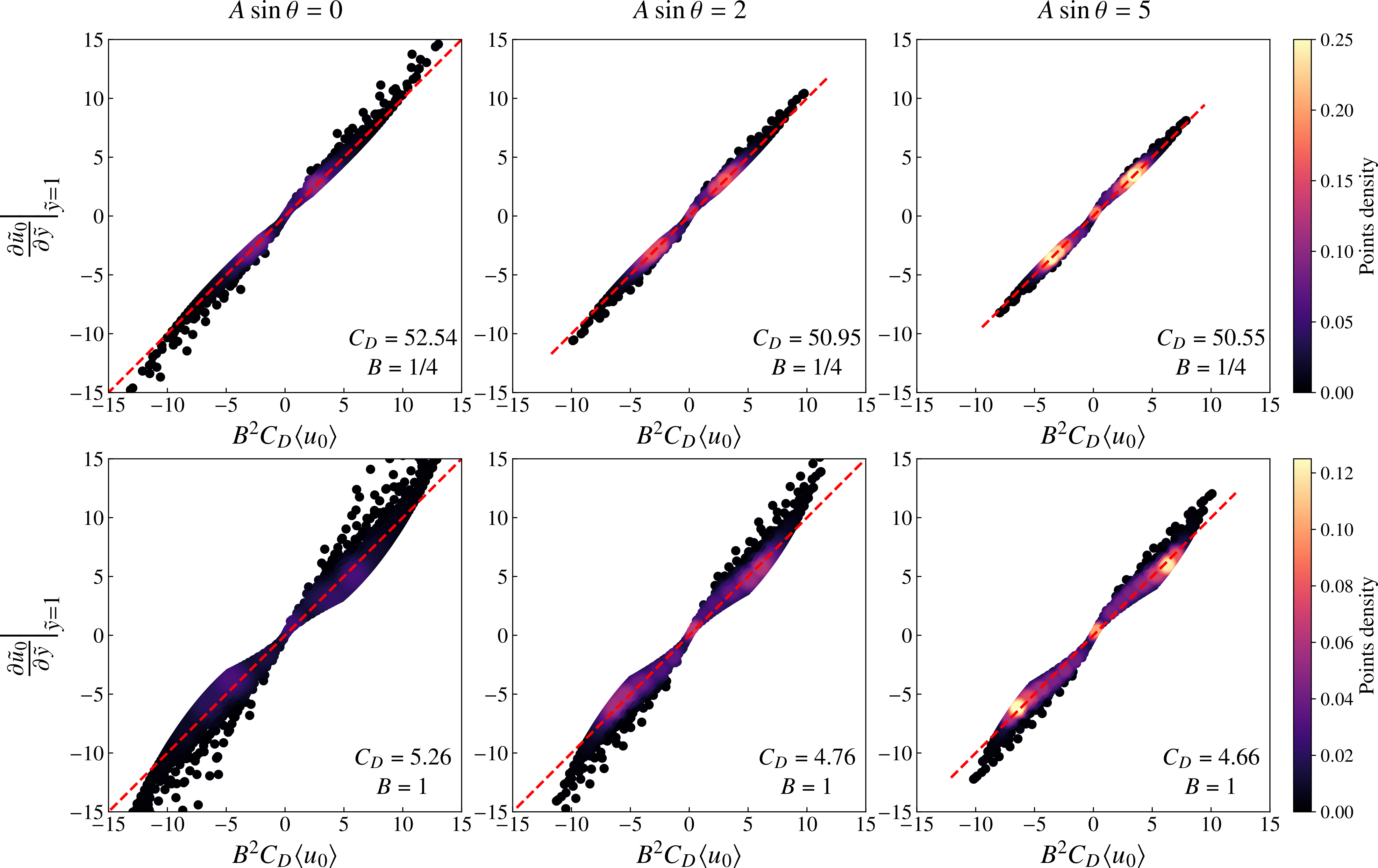}
    \caption{Empirical linear approximation of the sidewall-drag obtained from the 3D semi-analytical solution. Scatter plots compare the wall-shear term $B^{-2}\partial \tilde{u}_0/\partial \tilde{y}|_{\tilde{y}=1}$ with the width-averaged velocity $\av{u_0}$. The top and bottom rows correspond to $B=1/4$ and $B=1$, respectively. Columns show $A\sin\theta=0,2,5$, increasing from left to right. Colour denotes point density. Red dashed lines show least-squares linear fits, whose slopes define the effective sidewall drag coefficient $C_D$.}
    \label{fig:SidewallCoefficients}
\end{figure}

We determine $C_D$ from the semi-analytical solution by comparing the two sides of \eqref{eq:parametrisation_CD}. The relationship is not exactly linear, but the scatter plots in figure~\ref{fig:SidewallCoefficients} show that a linear drag law gives a good approximation over the range of $B$ and $A\sin\theta$ considered. We therefore estimate $C_D$ from a least-squares fit. The linear approximation is strongest for narrow ducts, where sidewall effects are most important: for $B=1/4$, the fits have $R^2>0.99$. For $B=1$, the scatter is slightly larger, but the fit remains reasonable, with $R^2>0.94$ in the cases shown.

\begin{figure}
    \centering \includegraphics[width=0.5\textwidth]{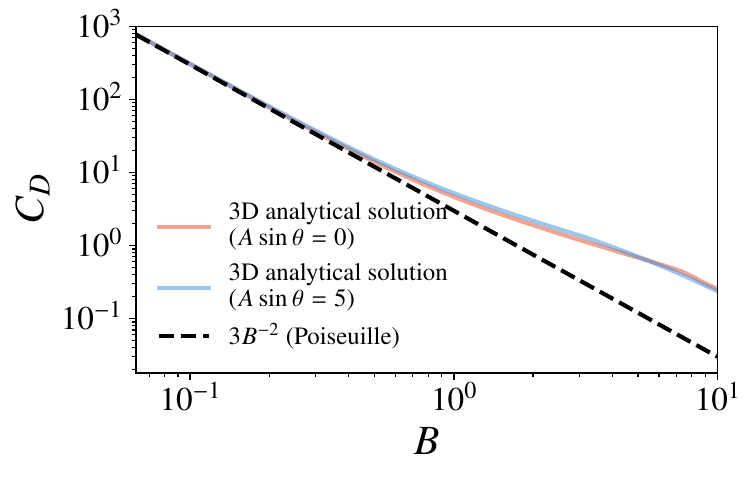}
    \caption{Effective sidewall drag coefficient $C_D$ as a function of the spanwise aspect ratio $B$. Values are obtained from least-squares fits of \eqref{eq:parametrisation_CD}, as illustrated in figure~\ref{fig:SidewallCoefficients}, for $A\sin\theta=0$ and $5$. The dashed line shows the narrow-duct Poiseuille scaling $C_D=3B^{-2}$.}
    \label{fig:CD_Asinth_B}
\end{figure}

Figure~\ref{fig:CD_Asinth_B} shows that the fitted values of $C_D$ depend primarily on the spanwise aspect ratio $B$, increasing rapidly as the duct narrows. By contrast, the dependence on $A\sin\theta$ is weak in the range $0\lesssim A\sin\theta\lesssim5$, with the curves for $A\sin\theta=0$ and $5$ nearly overlapping.
This dependence on $B$ is consistent with the spanwise velocity profiles in figure~\ref{fig:profileyB}. In the narrow-duct limit, the profile approaches a Poiseuille flow, for which $C_D=3B^{-2}$. This scaling, shown as a dashed line in figure~\ref{fig:CD_Asinth_B}, captures the fitted values well for $B\lesssim1$. However, in the calculations below we use the fitted $C_D$ obtained from the 3D semi-analytical solution, thereby retaining the weaker dependence on $A\sin\theta$.

\subsection{Width-averaged equations}
\label{sec:width-average}
We now formulate the time-dependent width-averaged equations used for the stability analysis and nonlinear simulations. For this purpose, we use the buoyancy velocity $U_g$ and the duct height $H$ to nondimensionalise the governing equations, matching the numerical formulation introduced in \S\ref{sec:numerics}. Dimensionless variables are denoted by a check accent and defined by
\begin{equation*}
    \vec{u} = U_g \check{\vec{u}},\quad  \vec{x}=\dfrac{H}{2}\check{\vec{x}},\quad \rho'=\frac{\Delta\rho}{2}\check{\rho}', \quad p=\bar{\rho}U_g^2\check{p}, \quad t=\frac{H}{2U_g}\check{t},
\end{equation*}
such that \eqref{eq:NS-before} becomes
\begin{subequations}
\label{eq:NonDimExperiments}
\begin{align}
    \frac{\partial \check{\vec{u}}}{\partial \check{t}} 
    + \check{\vec{u}} \cdot \check{\vec{\nabla}} \check{\vec{u}} 
    &= -\check{\vec{\nabla}} \check{p}
    + \frac{1}{\Rey_g} \check{\nabla}^2 \check{\vec{u}}
    + \frac{1}{4} \check{\rho}' \hat{\vec{g}},
    \label{eq:MomentumNonDimExperiments} \\[1.0ex]
    \check{\vec{\nabla}} \cdot \check{\vec{u}}
    &= 0,
    \label{eq:ContinuityNonDimExperiments} \\[1.0ex]
    \frac{\partial \check{\rho}'}{\partial \check{t}}
    + \check{\vec{u}} \cdot \check{\vec{\nabla}} \check{\rho}'
    &= \frac{1}{\Rey_g Sc} \check{\nabla}^2 \check{\rho}',
    \label{eq:AdvDifNonDimExperiments}
\end{align}
\end{subequations}
where $\hat{\vec{g}}=(\sin\theta,\ 0,\ -\cos\theta)$ is the unit vector in the direction of gravity, inclined at an angle $\theta$ to the duct-normal direction $\check{z}$ (figure~\ref{fig:setup}). The new scaled streamwise velocity $\check{u}$ is related to the dimensionless velocity $\tilde{u}$ used in the long-duct asymptotic analysis by $\check{u}=\Fr\,\tilde{u}$. Therefore, in the laminar regime described by the semi-analytical solution presented in \S\ref{sec:3Dsolution}, $\check{u}=\Fr^*\,\tilde{u}_0(A^{-1}\check{x},B^{-1}\check{y},\check{z}),
$ where $\tilde{u}_0$ is given by \eqref{eq:EquationForuwithn}.

To represent sidewall friction in a 2D framework, we introduce the width-averaged velocity
\begin{equation}\label{eq:width_averaged_velocity}
    \av{\check{\vec{u}}}=(\av{\check{u}},0,\av{\check{w}})=\dfrac{1}{2B}\int_{-B}^B(\check{u},\check{v},\check{w}) \, \mathrm{d}\check{y}.
\end{equation}
The boundary conditions give $\av{\check{v}}=0$. We also retain the assumption used in the base-flow model that the density field is independent of the spanwise coordinate, so that $\av{\check{\rho}'}=\check{\rho}'$. The scalar field is therefore governed by a 2D advection--diffusion equation. The width-averaged governing equations are
\begin{subequations}
\label{eq:width_averaged_system}
\begin{align}
\frac{\partial{\av{\check{\vec{u}}}}}{\partial{\check{t}}} 
+ \av{\check{\vec{u}}} \cdot\check{\vec{\nabla}}_{2D}\av{\check{\vec{u}}}
&= -\check{\vec{\nabla}}_{2D} \check{p}
+\frac{1}{\Rey_g}\check{\nabla}^2_{2D}\av{\check{\vec{u}}}
+\dfrac{1}{4}\check{\rho}' \hat{\vec{g}}
-\dfrac{1}{\Rey_g B}
\left. \pder{}{\check{\vec{u}}}{\check{y}}\right|_{\check{y}=B},
\label{eq:NonDinNVAvg} \\
\check{\vec{\nabla}}_{2D}\cdot\av{\check{\vec{u}}} 
&= 0,
\label{eq:ContNonDimWithSidewall}\\
\frac{\partial \check{\rho}'}{\partial \check{t}}
+ \av{\check{\vec{u}}} \cdot \check{\vec{\nabla}}_{2D} \check{\rho}' 
&= \frac{1}{\Rey_g Sc}\check{\nabla}^2_{2D}\check{\rho}'
\label{eq:AdvDiffNonDimWithSidewall}
\end{align}
\end{subequations}
with $\check{\vec{\nabla}}_{2D}=(\partial/\partial \check{x},\partial/\partial \check{z})$.

The sidewall contribution in \eqref{eq:NonDinNVAvg} is the time-dependent analogue of the wall-shear term parametrised in \eqref{eq:parametrisation_CD}. Using the effective drag coefficient $C_D$, we therefore write the width-averaged momentum equation as
\begin{equation}
    \frac{\partial{\av{\check{\vec{u}}}}}{\partial{\check{t}}} + \av{\check{\vec{u}}} \cdot\check{\vec{\nabla}}_{2D}\av{\check{\vec{u}}} = -\check{\vec{\nabla}}_{2D} \check{p} + \frac{1}{\Rey_g}\check{\vec{\nabla}}_{2D}^2\av{\check{\vec{u}}} 
    + \frac{1}{4}\check{\rho}' \hat{\vec{g}} 
    - \frac{C_D}{\Rey_g}(\av{\check{\vec{u}}}\cdot \hat{\vec{x}})\hat{\vec{x}},\label{eq:MomentumNonDimWithSidewall}
\end{equation}
where $\hat{\vec{x}}$ represents the unit vector in the streamwise direction. Equations~\eqref{eq:MomentumNonDimWithSidewall}, \eqref{eq:ContNonDimWithSidewall}, and \eqref{eq:AdvDiffNonDimWithSidewall} define the 2D width-averaged model used below. The effect of lateral confinement enters only through the fitted drag coefficient $C_D(B,A\sin\theta)$. This formulation preserves the base-flow structure predicted by the 3D semi-analytical solution while reducing the stability problem to 2D. We now use it to determine when the laminar exchange flow first becomes unstable to Holmboe waves, as a leading-order representation of sidewall friction while assuming that the spanwise velocity profile remains close to that of the base flow. 

\subsection{Linear stability analysis}
\label{sec:LSA}

\subsubsection{Formulation}
We use a temporal linear-stability formulation with 2D perturbations about a one-dimensional base flow, following the 2P--1B framework of \citet{Lefauve2018} and \citet{Ducimetiere2021EffectsInstabilities}. Unlike these previous studies, our base flow is not prescribed independently: it is taken from the width-averaged semi-analytical SID solution derived above, and sidewall effects enter through the effective drag coefficient $C_D$. The aim is to determine the threshold for interfacial-wave growth in the laminar width-averaged flow as a function of $\Rey_g$, $A\sin\theta$, and $B$.

The base velocity is taken to be a one-dimensional parallel flow,
$\vec{U}_b\equiv[U_b(\check{z}),0,0]$, and the corresponding base density is denoted by $R(\check{z})$, defined as
\begin{equation}
    U_b(\check{z})
    =
    \Fr^*
    \left\langle
    \tilde{u}_0(A^{-1}\check{x},B^{-1}\check{y},\check{z})
    \right\rangle_{\check{x}=\check{x}_E},
    \qquad
    R(\check{z})
    =
    \tanh{\left(\frac{\check{z}-\eta(A^{-1}\check{x}_E)}{\delta_R}\right)}.
\end{equation}
Here $\check{x}_E$ is the streamwise position at which the local stability problem is evaluated, and $\delta_R=0.05$. Hereafter, we present this position through the rescaled coordinate $\tilde{x}_E\equiv A^{-1}\check{x}_E\in[-1,1]$, so that $\tilde{x}_E=\pm1$ corresponds to the duct ends. The smoothed density profile $R$ replaces the sharp two-layer interface in \eqref{eq:density}, providing the continuous base density required for the LSA and consistent with values used in previous studies \citep{Lefauve2018,Ducimetiere2021EffectsInstabilities}. The results are weakly sensitive to the exact value of $\delta_R$ provided $\delta_R\ll1$ and the gradient is adequately resolved. Evaluating both profiles at a given streamwise position effectively gives a local parallel-flow approximation, which should hold well in long ducts because the interface inclination is small, except near the ends where the interface slope increases. For this reason, we restrict the analysis to $|\tilde{x}_E|\leq0.7$.

The base flow is subjected to small perturbations of the form 
\begin{equation}
\vec{\check{u}} = [U_b(\check{z})+\xi \check{u}^*(\check{x},\check{z},\check{t})]\hat{\vec{x}}+\xi \check{w}^*(\check{x},\check{z},\check{t})\hat{\vec{z}}, \quad \text{and}\quad \check{\rho}' = R(\check{z})+\xi \check{\rho}^*(\check{x},\check{z},\check{t})
\end{equation}
with $\xi\ll1$. 
The perturbations are written as normal modes, with the real part implied,
$\check{\psi}^*(\check{x},\check{z},\check{t})=\check{\psi}(\check{z})\exp(ik\check{x}+\sigma \check{t})$, where $\check{\psi}(\check{z})\in\mathbb{C}$ is the eigenfunction, $k>0$ is the streamwise wavenumber, and $\sigma\in\mathbb{C}$ is the temporal eigenvalue. The real part of $\sigma$ is the perturbation growth rate, while its imaginary part is the angular frequency.

Eliminating the streamwise velocity and pressure perturbations using continuity gives the generalised eigenvalue problem
\begin{equation}
\sigma
\begin{bmatrix}
\check{\nabla}_{2D}^2 & 0\\
0 & \mathcal{I}
\end{bmatrix}
\begin{bmatrix}
\check{w}\\
\check{\rho}
\end{bmatrix}
=
\begin{bmatrix}
\mathcal{L}_{\check{w}} & \mathcal{L}_{\check{w}\check{\rho}}\\
\mathcal{L}_{\check{\rho}\check{w}} & \mathcal{L}_{\check{\rho}}
\end{bmatrix}
\begin{bmatrix}
\check{w}\\
\check{\rho}
\end{bmatrix},
\label{eq:LSA_EVP}
\end{equation}
where $\mathcal{I}$ denotes the identity operator, and the operators on the right hand side are
\begin{subequations}
\begin{align}
\mathcal{L}_{\check{w}} 
&\equiv 
-ikU_b\check{\nabla}_{2D}^2 
+ ik\pder{2}{U_b}{\check{z}}
+ \frac{1}{\Rey_g}\check{\nabla}_{2D}^4 
- \underbrace{\frac{C_D}{\Rey_g}\check{\nabla}_{2D}^2}_{\text{sidewall friction}}, 
\label{eq:lw}\\
\mathcal{L}_{\check{\rho}} 
&\equiv 
-ikU_b 
+ \frac{1}{\Rey_g Sc}\check{\nabla}_{2D}^2,\\
\mathcal{L}_{\check{w}\check{\rho}} 
&\equiv 
\frac{1}{4}
\left(k^2\cos\theta - ik\sin\theta\,\pder{}{}{\check{z}}\right),\\
\mathcal{L}_{\check{\rho}\check{w}} 
&\equiv 
-\pder{}{R}{\check{z}}
\end{align}
\end{subequations}
with $\check{\nabla}_{2D}^2= -k^2+\partial^2/\partial\check{z}^2
$ and $
\check{\nabla}_{2D}^4= k^4+\partial^4/\partial\check{z}^4
-2k^2\partial^2/\partial\check{z}^2$.
In \eqref{eq:lw}, the last 2D Laplacian with a factor $C_D/\Rey_g$ represents the parametrised effect of sidewall friction on $\mathcal{L}_{\check{w}}$, which is novel compared to the classical 2P--1B formulations of \citet{Lefauve2018} and \citet{Ducimetiere2021EffectsInstabilities}. It acts as an effective Rayleigh damping on the velocity perturbation and reduces growth rates by $\approx C_D/\Rey_g$, but it is not an exact uniform shift of all eigenvalues because it does not act directly on the density perturbation. 

Vertical confinement is represented by no-slip and no-penetration conditions for the velocity perturbation, which in the present formulation give $\check{w}=\partial_{\check{z}}\check{w}=0$, and by a no-flux condition for the density perturbation, $\partial_{\check{z}}\check{\rho}=0$, at $\check{z}=\pm1$.

The formulation is therefore the classical 2P--1B stability problem, modified only by the sidewall-damping term in the perturbation momentum equation. The semi-analytical base flow further allows us to evaluate stability over a broad range of $(\Rey_g , A\sin\theta, B)$, while retaining the effect of confinement on both the base flow and perturbation damping. This formulation differs from the confined stability analysis of \citet{Ducimetiere2021EffectsInstabilities}, where sidewall effects were represented by prescribing a Poiseuille-like spanwise-dependent base flow and allowing the perturbation modes to vary across the duct width. Here, the stability problem is 2D and width-averaged, allowing the linear stability computations and nonlinear simulations below to explore a multi-dimensional parameter space.

\subsubsection{Results}
\label{sec:LSAResults}

We use LSA to determine the onset of Holmboe waves. For a given base flow, we consider the flow unstable if at least one mode in the range $0.1\leq k \leq 5$ satisfies $\Re[\sigma(k)] > 0$. For each set of imposed parameters, we then determine the critical value of the laminar Froude number $\Fr^*_c$ at which this condition is first satisfied for different values of the spanwise aspect ratio $B$, the geometric parameter $A\sin\theta$, and the streamwise position $\tilde{x}_{E}$. The Reynolds number is given by \eqref{eq:Fr*}, so $\Rey_g=A K \Fr^*$ where $K$ is determined for given values of $B$ and $A\sin \theta$ as described in \S\ref{sec:interface}. We consider here $A=30$, which was tested to be large enough. The governing geometric parameters $A \sin \theta$ and $B$ influence the eigenvalue problem through two related effects: they determine the value of the parameter $K$ (affecting the width-averaged velocity field) and set the strength of the effective sidewall friction through the drag coefficient $C_D$.

\begin{figure}
    \centering
    \includegraphics[width=0.8\linewidth]{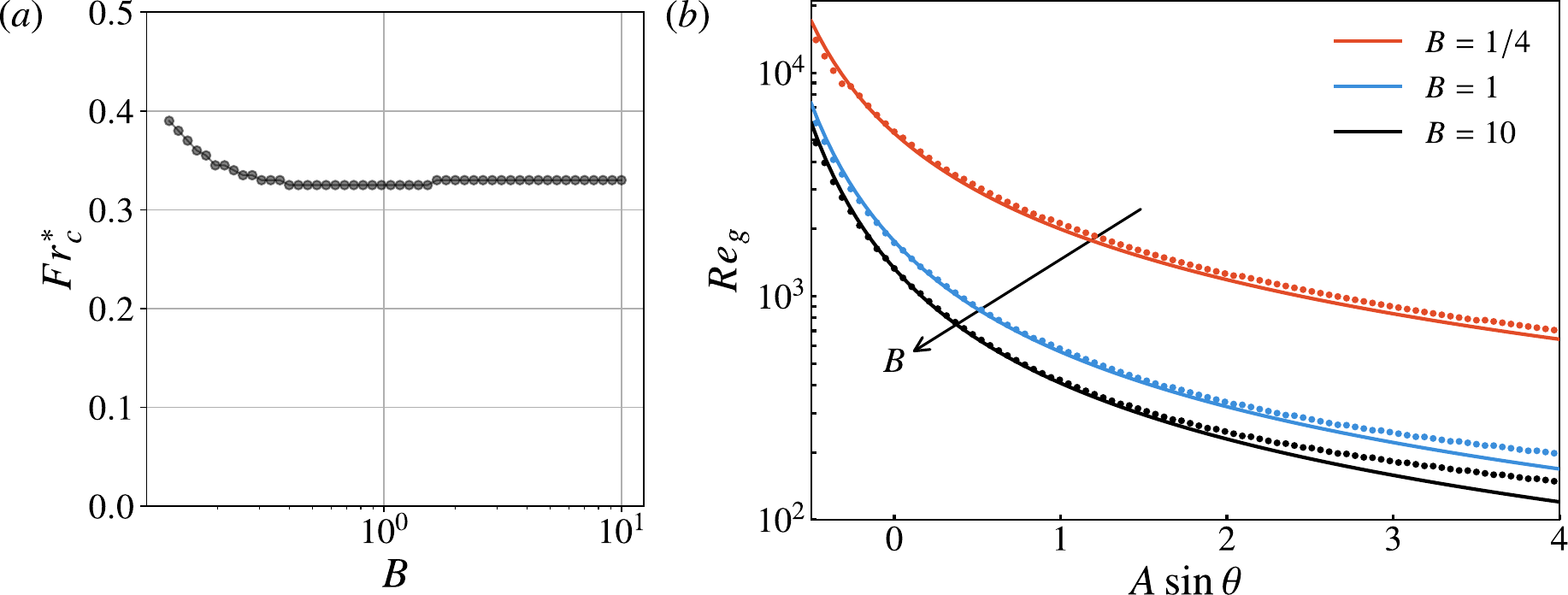}
    \caption{Influence of the spanwise aspect ratio $B$ and the geometrical parameter $A\sin\theta$ on the flow instability. (\textit{a}) The effect of $B$ on the critical value $\Fr^*_c$ at $\tilde{x}_E=0$ for the case with $A \sin\theta=0$. (\textit{b}) Stability map from the LSA in the ($\Rey_g, A\sin\theta$) parameter space for three distinct spanwise aspect ratios $B=[1/4,\ 1,\ 10]$ and $A=30$. The dotted curves represent the value of $\Fr^*_c$ as obtained from the LSA. The solid curves represent constant values $\Fr^*=0.33$ for the three values of $B$. The flow is stable below the curves, and unstable above them.}
    \label{fig:LSAresults1}
\end{figure}

We first consider the stability of the flow at $\tilde{x}_E=0$. Figure~\ref{fig:LSAresults1}(\textit{a}) shows $\Fr^*_c$ as a function of $B$ for $A\sin\theta=0$. The LSA was performed for $0.2\leq\Fr^*\leq0.4$ with increments of $0.005$. For sufficiently wide ducts, $\Fr^*_c$ is nearly independent of $B$ and approaches the 2D value $\Fr^*_c\simeq0.33$. In this regime, the width-averaged velocity profile varies only weakly with $B$ and the sidewall damping $C_D/\Rey_g$ is too small to significantly affect the instability threshold.

For narrow ducts, $B\lesssim0.25$, the critical Froude number increases sharply. This stabilisation has two related causes: sidewall friction increases through $C_D$ (recall figure~\ref{fig:CD_Asinth_B}), and the laminar base flow develops weaker interfacial shear (recall figure~\ref{fig:profileyB}(\textit{b})). In the limit $B\to0$, $C_D\to\infty$, and interfacial-wave growth is suppressed. Thus, lateral confinement affects instability through two distinct mechanisms: indirectly by modifying the laminar base flow and directly by damping perturbations.

Next, we examine how the stability threshold varies with $A\sin\theta$. Figure~\ref{fig:LSAresults1}(\textit{b}) shows the critical value $\Fr^*_c$ at $\tilde{x}_E=0$ for $B=[1/4, 1,10]$, plotted in the $(\Rey_g,A\sin\theta)$ parameter space. The coloured markers indicate the first value of $\Fr^*$ for which at least one mode is unstable. The solid, dashed, and dotted curves show constant-$\Fr^*=\Fr^*_c$ contours for the same values of $B$. For all three values of $B$, the instability threshold remains close to $\Fr^*_c\simeq0.33$. The weak dependence on $A\sin\theta$ follows from the local base flow at $\tilde{x}_E=0$: since $\eta(0)=0$, the central velocity profile depends primarily on $B$, while $A\sin\theta$ enters only weakly through $C_D$ and the small tilt-dependent terms in the stability operator. Thus, near the duct centre, the onset of interfacial waves is governed by an approximately constant critical value of the confinement-adjusted Froude number $\Fr^*$ throughout the range of $A\sin\theta$ considered. We compare this prediction with the numerical and experimental results in \S\ref{sec:numericalResults}.

\begin{figure}
    \centering
    \includegraphics[width=0.45\linewidth]{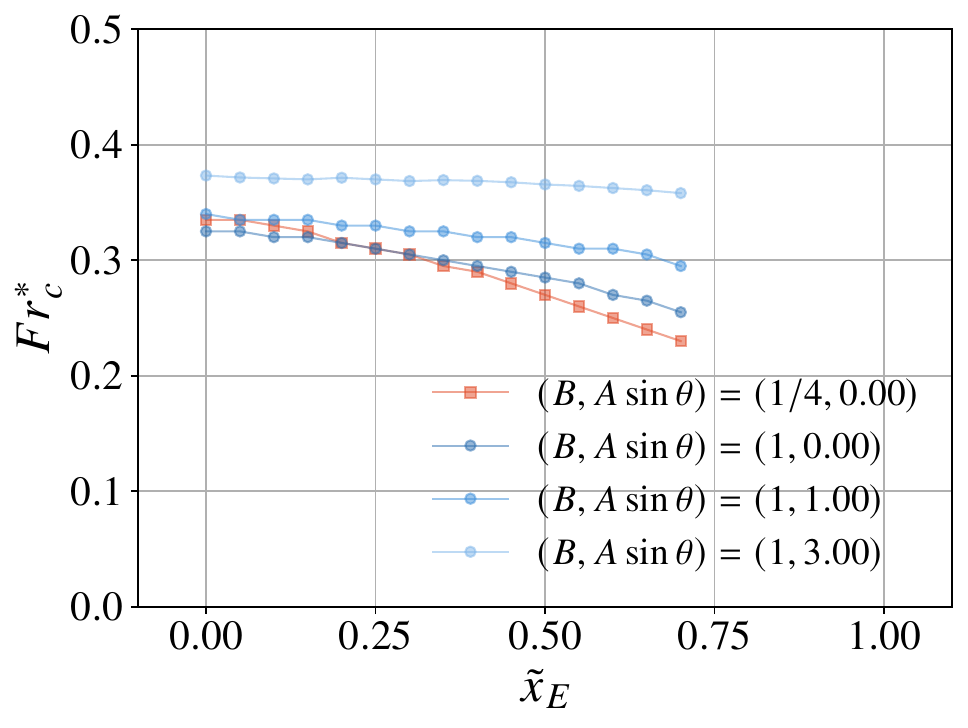}
    \caption{Effect of along-duct position on the linear instability threshold. The critical laminar Froude number $\Fr^*_c$ is shown as a function of the evaluation position  $\tilde{x}_E$ for four combinations of $B$ and $A\sin\theta$. The figure illustrates that the central threshold at $\tilde{x}_E=0$ is not necessarily representative along the full duct, especially in narrow, horizontal ducts (orange curve). The variation of $\Fr^*_c$ reflects changes in the local interface slope, layer thickness, and sidewall damping. The analysis is restricted to $|\tilde{x}_E|\leq 0.7$, because the local parallel-flow approximation becomes less accurate near the duct ends.}
    \label{fig:LSAresults2}
\end{figure}

Finally, we examine how the stability threshold varies with the streamwise evaluation position $\tilde{x}_E$. Figure~\ref{fig:LSAresults2} shows $\Fr^*_c$ as a function of $\tilde{x}_E$ for four representative cases: three cases with $B=1$ and different values of $A\sin\theta$, and one narrower case with $B=1/4$ and $A\sin\theta=0$. At $\tilde{x}_E=0$, the three cases with $B=1$ nearly coincide, confirming the weak central dependence on $A\sin\theta$. By contrast, the $B=1/4$ case has a larger $\Fr^*_c$, consistent with stronger sidewall damping. Away from the duct centre, $\Fr^*_c$ depends more strongly on $\tilde{x}_E$. As $|\tilde{x}_E|$ increases, one of the two layers becomes thinner, changing the interfacial shear of the local base flow. For cases with a flatter interface around $\tilde{x}=0$, $\Fr^*_c$ is more uniform along the duct, while cases with a steeper interface show a decrease in $\Fr^*_c$ toward the duct ends. This trend is consistent with the interface shapes shown in figure~\ref{fig:eta}.

The LSA results clarify the interpretation of the stability threshold. Near $\tilde{x}_E=0$, the onset of interfacial waves is well represented by an approximately constant $\Fr^*_c\simeq0.33$ for sufficiently wide ducts, $B\gtrsim0.25$. Away from the duct centre, this threshold is lower, consistent with experimental observations that interfacial waves appear first closer to the duct ends \citep{Lefauve2018thesis}. However, these edge values should be interpreted with caution because the local parallel-flow approximation becomes less accurate as $|\tilde{x}_E|\to1$. The robust conclusion is that, if the flow is unstable at the duct centre, it is unstable throughout the duct.

\section{Width-averaged simulations and regime transitions} \label{sec:numericalResults}

\subsection{Numerical setup} \label{sec:numerics}

We now use the width-averaged formulation of \S\ref{sec:width-average} to perform 2D time-dependent simulations. These simulations have two purposes. First, they test whether the LSA threshold predicts the onset of waves in an unsteady, fully nonlinear flow. Second, they allow direct comparison with experimental regime diagrams over a wide range of $\Rey_g$, $A\sin\theta$, and $B$. Because sidewall effects enter through the effective drag coefficient $C_D$, the model retains the leading influence of lateral confinement at a much lower computational cost than fully 3D simulations \citep[e.g.][]{Zhu2023}.

\begin{figure}
    \includegraphics[width=1\textwidth]{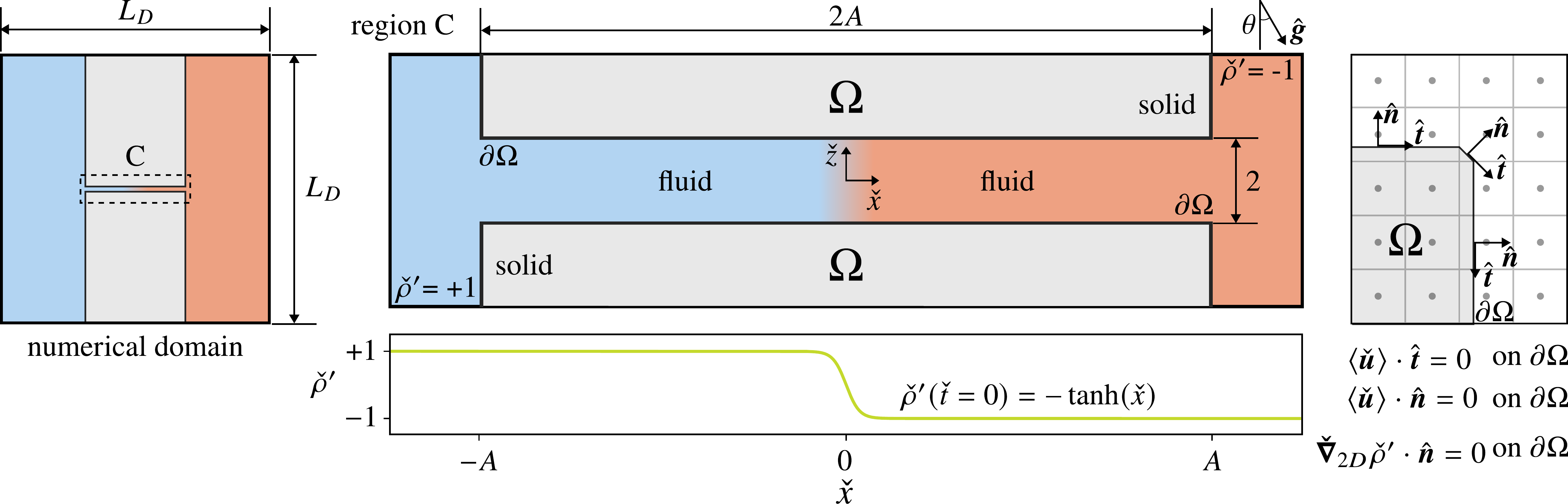}
    \caption{Numerical setup for the width-averaged simulations. The  domain is a square of side length $L_D$, containing two artificial reservoirs connected by a central duct of length $2A$ and height $2$. The fluid region is bounded by an embedded solid $\Omega$, whose boundary $\partial\Omega$ cuts the Cartesian mesh cells. Gravity acts at an angle $\theta$ relative to the duct-normal direction, providing forcing to the exchange flow. The simulations are initialised with the lock-exchange density field $\check{\rho}'(\check{t}=0)=-\tanh(\check{x})$, giving dense fluid on one side and light fluid on the other. On $\partial\Omega$, the width-averaged velocity satisfies no-slip and no-penetration conditions, $\av{\check{\vec{u}}}\cdot\hat{\vec{t}}=0$ and $\av{\check{\vec{u}}}\cdot\hat{\vec{n}}=0$, while the density satisfies the no-flux condition $\vec{\check{\nabla}}_{2D}\check{\rho}'\cdot\hat{\vec{n}}=0$.}
    \label{fig:SchematicNumerics}
\end{figure}

We solve the dimensionless width-averaged governing equations \eqref{eq:ContNonDimWithSidewall}, \eqref{eq:AdvDiffNonDimWithSidewall}, and \eqref{eq:MomentumNonDimWithSidewall} using the open-source solver \textit{Basilisk} \citep{Popinet2009,Popinet2015}. The computational domain is a square of side length $L_D=160$, containing two reservoirs connected by a central duct of dimensions $2A\times 2$, as shown in figure~\ref{fig:SchematicNumerics}. The duct walls and reservoirs are represented using Basilisk's embedded boundary method (EBM), in which the solid region $\Omega$ cuts the underlying Cartesian mesh. No-slip and no-penetration conditions are imposed on the embedded boundary, $\av{\check{\vec{u}}}\cdot\hat{\vec{t}}=0$ and $\av{\check{\vec{u}}}\cdot\hat{\vec{n}}=0$, where $\hat{\vec{n}}$ and $\hat{\vec{t}}$ are the unit normal and tangent to $\partial\Omega$. For density, we impose a no-flux condition, $\check{\vec{\nabla}}_{2D}\check{\rho}'\cdot\hat{\vec{n}}=0$, on the solid boundaries. The same boundary conditions are applied at the outer boundaries of the computational domain. The reservoirs are chosen sufficiently large that they do not affect the flow in the central duct over the simulated time interval. The EBM in Basilisk has been extensively validated for different physical problems \citep[\textit{e.g.}][]{Xue2023,Konstantinos2024}. For further details on the algorithm used in the EBM implementation in Basilisk, the reader is referred to \citet{Johansen1998} and \citet{Schwartz2006}.

The simulations use a second-order, cell-centred finite-volume scheme on a Cartesian grid. Adaptive mesh refinement is based on wavelet estimates of the discretisation error, allowing refinement in regions with strong velocity or density gradients. The refinement tolerance is set to $10^{-3}$ for both $\check{\rho}'$ and $\av{\check{\vec{u}}}$. The minimum and maximum grid spacings are $\Delta_{\min,\max}=L_D/2^{n_{\min,\max}}$, where $n_{\min}$ and $n_{\max}$ are the coarsest and finest refinement levels. To reduce computational cost, refinement is relaxed in regions far from the duct, where fine resolution is unnecessary. Grid-independence tests were performed on a range of refinement levels showing that $n_{\min} = 8$ and $n_{\max} = 13$ were sufficient. A generalised minmod flux limiter \citep{VanLeer1979} is used to prevent non-physical density overshoots beyond $\check{\rho}'=\pm1$.

We consider three geometries, denoted wSID, mSID, and tSID, corresponding to the wide, mini, and tall experimental configurations of \citet{Lefauve2020} and \citet{Lefauve2024}. Their parameters are listed in Table~\ref{tab:ExpDetails}. The simulations are initialised from a lock-exchange density field \citep{Wood1970,Hartel2000,Shin2004}, $\check{\rho}'(\check{t}=0)=-\tanh(\check{x})$, as shown in figure~\ref{fig:SchematicNumerics}. This initial condition creates a sharp density gradient at $\check{x}=0$. To remove the influence of initial transients, all analysis is performed only after the flow has reached a statistically established state, typically for $\check{t}\gtrsim100$.

\begin{table*}
\centering

\renewcommand{\arraystretch}{1.3}
\begin{tabularx}{\textwidth}{XX>{\centering\arraybackslash}m{2cm}XXXX>{\hsize=1.05\hsize}X}
\hline
Case & $H$ (mm) & Cross-section & $A$ & $B$ & $\theta$ (deg.) & $\Rey_g$ \\
\hline
wSID & 50 & \tikz{\draw[thick] (0,0) rectangle (1,0.5);} & 40 & 2 & $[0,6]$ & $[300, 6000]$ \\
mSID & 45 & \tikz{\draw[thick] (0,0) rectangle (0.5,0.5);} & 30 & 1 & $[-1,6]$ & $[300, 6000]$ \\
tSID & 90 & \tikz{\draw[thick] (0,0) rectangle (0.25,1);} & 15 & 1/4 & $[-1,3]$ & $[3000, 15000]$ \\
\hline
\end{tabularx}
\caption{Parameters of the experimental configurations reproduced by the width-averaged simulations. The three geometries are wSID \citep{Lefauve2024}, mSID, and tSID \citep{Lefauve2020}. The table lists the duct height $H$, cross-section, longitudinal aspect ratio $A$, spanwise aspect ratio $B$, imposed tilt-angle range $\theta$, and buoyancy Reynolds-number range $\Rey_g$. All cases correspond to salt stratification in water, with $Sc\simeq700$.}
\label{tab:ExpDetails}
\end{table*}

\subsection{Validation of the width-averaged simulations}

We first validate the width-averaged simulations against the semi-analytical laminar solution of \S\ref{sec:3Dsolution}. The comparison, reported in Appendix~\ref{sec:AppendixAnalyticalComp}, is good for both the interface shape and the velocity field. The simulations also reproduce the effect of lateral confinement: for smaller $B$ values, the exchange velocity is reduced and the vertical velocity profiles are modified, while for larger $B$ values, the results approach the weakly confined limit. This agreement confirms that the numerical model recovers the laminar base flow from which interfacial waves emerge.

We next compare the simulations with the experimental regime classifications of \citet{Lefauve2020} and \citet{Lefauve2024}, using the datasets made available by \citet{Lefauve2020a} ($B=1$ and $1/4$) and \citet{Jiang2023} ($B=2$). Following \citet{Macagno1961} and \citet{Meyer2014}, we distinguish four regimes: laminar flow, interfacial waves, intermittently turbulent flow, and fully turbulent flow. In the SID experiments considered here, all cases use salt stratification with $Sc\simeq700$, and the observed interfacial waves are clearly identified as Holmboe waves.

A detailed comparison with three experimental cases is presented in Appendix~\ref{sec:AppendixExperimentalComp}. Representative snapshots of the simulated and experimental density fields are shown in figure~\ref{fig:ExperimentalComp}, illustrating that the width-averaged model reproduces the characteristic laminar, Holmboe-wave, and intermittently turbulent flow structures. This agreement supports the use of the width-averaged model to capture the onset of waves at a high Schmidt number, $Sc\simeq700$, a regime that is prohibitively expensive for fully 3D simulations over the present parameter range. Some discrepancies start to be observed in the intermittently turbulent regime, where the 2D simulations produce longer-lived and larger vortices than those observed experimentally. We therefore focus on the transition from laminar flow to Holmboe waves, while using the simulations only qualitatively to identify the subsequent transition towards turbulence. 

\subsection{Regime transitions and comparison with LSA}
\label{sec:results_num}

\begin{figure}
    \centering    
    \includegraphics[width=\linewidth]{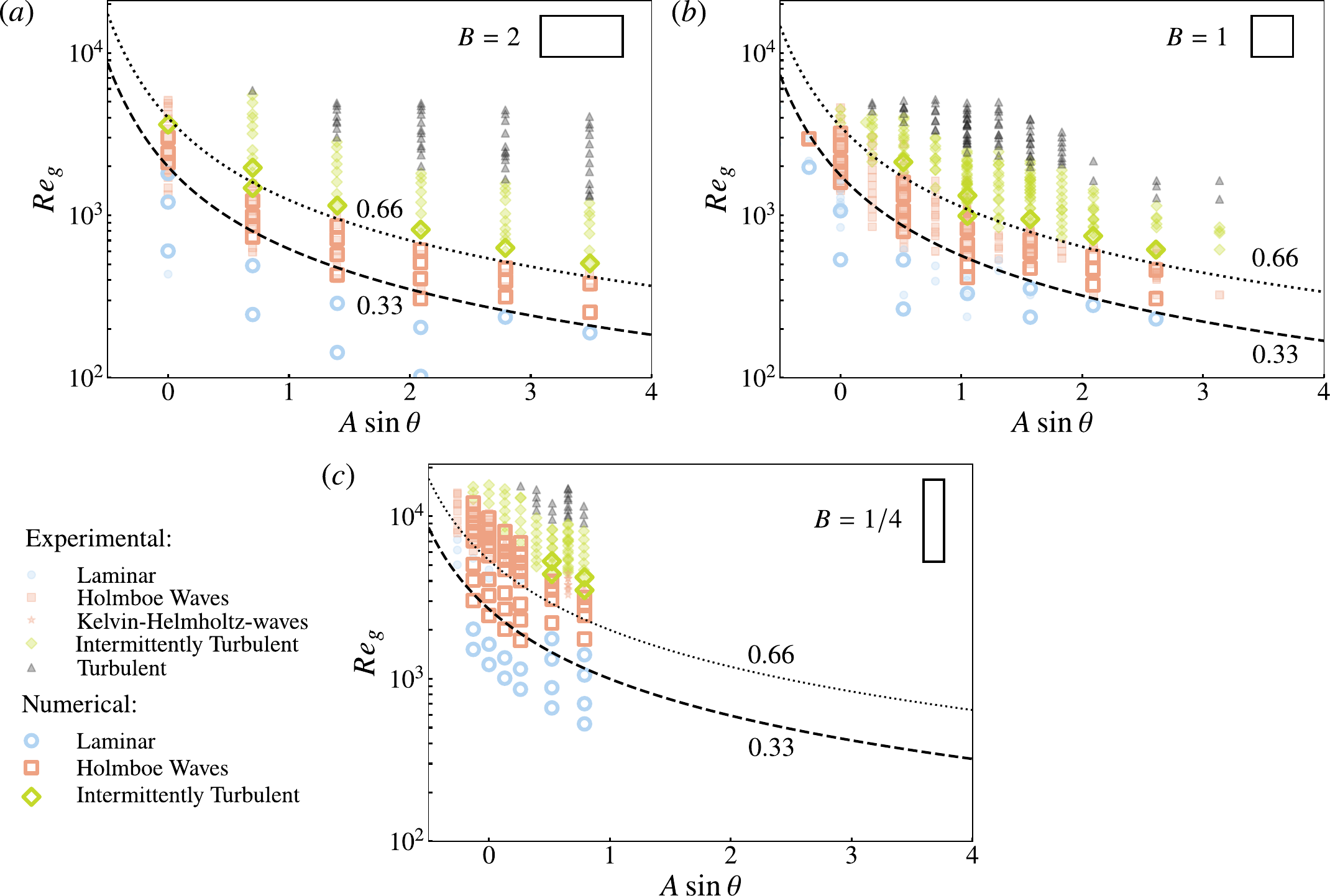}
\caption{Regime diagrams showing the collapse of the laminar-to-Holmboe-wave transition with the confinement-adjusted laminar Froude number $\Fr^*$. Results are shown for the three SID experimental configurations: (\textit{a}) wSID ($B=2$), (\textit{b}) mSID ($B=1$), and (\textit{c}) tSID ($B=1/4$). Experimental data are shown by filled symbols and width-averaged simulations by open symbols. The dashed line, $\Fr^*=0.33$, marks the laminar-to-Holmboe-wave threshold predicted by the LSA and largely collapses a transition that appeared geometry-dependent in previous studies. The dotted line, $\Fr^*\approx0.66$, marks the approximate transition from Holmboe waves to intermittent turbulence, which also shows strong collapse for $B\gtrsim1$. Predicting this second threshold from first principles, however, remains an open problem.}
    \label{fig:RegimesMapping}
\end{figure}

Figure~\ref{fig:RegimesMapping} compares numerical and experimental regime classifications in the $(\Rey_g,A\sin\theta)$ parameter space for (\textit{a}) wSID, (\textit{b}) mSID, and (\textit{c}) tSID. Numerical simulations are shown by large open symbols, and laboratory experiments by small filled symbols. The regimes are classified according to the temporal evolution of the density field. Overall, the width-averaged simulations reproduce the experimentally observed laminar, Holmboe-wave, and intermittently turbulent regimes. For wSID and mSID, discrepancies are limited mainly to the transition boundaries. For tSID, the numerical simulations predict laminar flow at slightly lower $\Rey_g$ for a given $A\sin\theta$, but capture the remaining regimes well.

The main result of this paper is that the laminar-to-Holmboe-wave transition collapses across the numerical cases at $\Fr^*\simeq0.33$, in agreement with the LSA threshold and the experimental regime maps. This collapse shows that the confinement-adjusted laminar Froude number $\Fr^*$, calculated from the semi-analytical base flow, provides a predictive threshold for wave onset in moderate and wide ducts, and that the width-averaged numerical framework captures the essential transition physics in the experiments. 

The subsequent transition from Holmboe waves to intermittent turbulence occurs near $\Fr^*\simeq0.66$ for wSID and mSID, and at a slightly larger value for tSID. Interestingly, this suggests that the same confinement-adjusted Froude number also organises the subsequent transition to intermittent turbulence, although no theoretical justification currently exists for this observation. Whether this reflects a broader dynamical role of $\Fr^*$ remains an open question.
The approximate factor of two between the laminar-to-wave and wave-to-turbulence thresholds was previously noted by \citet{Duran-Matute2023}. The slightly larger threshold in the narrow tSID geometry may reflect the influence of confinement on the secondary instabilities and 3D hairpin vortices that emerge from Holmboe waves and ultimately lead to intermittent turbulence under increased forcing \citep{Jiang2022}. These strongly nonlinear dynamics are not expected to be captured quantitatively by the present width-averaged model, which is designed primarily to predict the laminar-to-wave transition. Moreover, the linear drag formulation adopted in the width-averaged model becomes increasingly inappropriate as the flow transitions towards turbulence.

Our results extend the framework of \citet{Duran-Matute2023}, who showed that SID regime transitions can be organised using a laminar Froude number computed from a 2D, infinitely wide theory. In that framework, different duct geometries gave different apparent critical values, indicating that lateral confinement was not fully accounted for. Here, $\Fr^*$ is computed from the 3D semi-analytical base flow, including the effect of sidewalls on the laminar exchange flow. For moderate and wide ducts ($B>1/4$), this confinement-adjusted $\Fr^*$ is sufficient to collapse the laminar-to-Holmboe-wave transition to $\Fr^*\simeq0.33$. However, the LSA suggests that for narrower ducts, the critical value still increases because the side walls also damp the perturbations directly through $C_D$. 

The regime classifications in figure~\ref{fig:RegimesMapping} are based on observations near the duct centre. In the simulations this corresponds to $\check{x}=0$, whereas the experimental visualisations typically sample a finite region away from the duct ends. However, the LSA predicts that waves may appear at lower $\Fr^*$ closer to the ends, where the local base flow is more strongly asymmetric. To examine where waves first emerge in the simulations, we compute the kinetic energy of the temporal velocity fluctuations,
\begin{equation}
    \langle k'\rangle_z
    =
    \frac{1}{4}\int_{-1}^{1}
    \left(\check{u}'^{\,2}+\check{w}'^{\,2}\right)
    \,\mathrm{d}\check{z},
    \label{eq:kprime_z}
\end{equation}
where $\langle\cdot\rangle_z$ denotes the average over the duct height, and $\check{u}'=\av{\check{u}}-\overline{\av{\check{u}}}$ and $\check{w}'=\av{\check{w}}-\overline{\av{\check{w}}}$ are the temporal fluctuations of the width-averaged velocity components about their local time averages, denoted by an overbar. The time average is computed after the initial transient, once the exchange flow is established ($\check{t}\geq100$).
 
\begin{figure}
    \centering
    \includegraphics[width=0.9\linewidth]{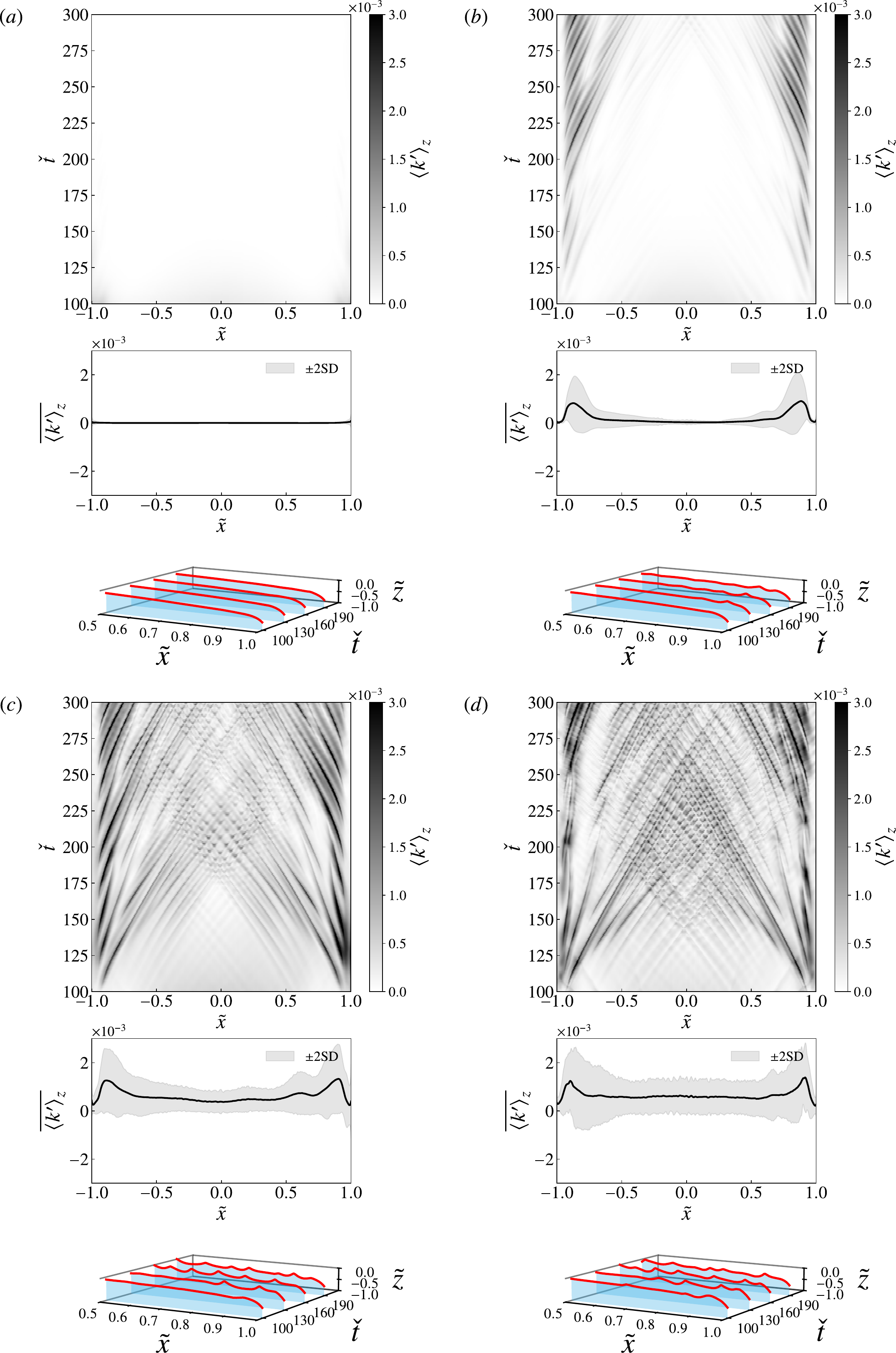}
    \caption{Development of wave-induced velocity fluctuations in four mSID simulations with $A\sin\theta=1.05$ and increasing forcing: $\Fr^*=[0.20,\,0.25,\,0.30,\,0.35]$ in (\textit{a})--(\textit{d}), respectively. For each case, the top row shows an $(\check{x},\check{t})$-diagram of the vertically averaged fluctuation energy $\langle k'\rangle_z$; the middle row shows its time average $\overline{\langle k'\rangle_z}$ as a function of $\check{x}$, with the shaded region indicating $\pm2\mathrm{SD}$; and the bottom row shows the interface position at $\check{t}=[100,130,160,190]$. For subcritical central values of $\Fr^*$, waves first appear near the duct ends and decay towards the centre; as $\Fr^*$ increases, they penetrate further into the duct interior.}
    \label{fig:fluctuations_changeFr}
\end{figure}

Figure~\ref{fig:fluctuations_changeFr} shows $\langle k'\rangle_z$ for four mSID simulations with $A\sin\theta=1.05$ and $\Fr^*=[0.20, 0.25,0.30,0.35]$ in panels (\textit{a}--\textit{d}), respectively. For each case, we show an ($\check{x},\check{t}$)-diagram of fluctuation energy, its time average $\overline{\langle k'\rangle_z}$ as a function of $\tilde{x}$, together with the corresponding standard deviation (SD), and the interface position at four times: $\check{t}=100$, $130$, $160$, and $190$. For $\Fr^*=0.20$, no appreciable fluctuations or interface deformations are observed. For $\Fr^*=0.25$, fluctuations and interface deformations appear only near the duct ends. For $\Fr^*=0.30$, waves are present over a larger fraction of the duct, but their energy decreases towards the centre, consistent with waves emerging near the ends and decaying as they propagate into a locally stable central region. For $\Fr^*=0.35$, waves are present throughout the duct, with nearly uniform fluctuation energy throughout the central region. These results agree with the LSA prediction that the flow is unstable throughout the duct once $\Fr^*\gtrsim0.33$.

\begin{figure}
    \centering
    \includegraphics[width=\linewidth]{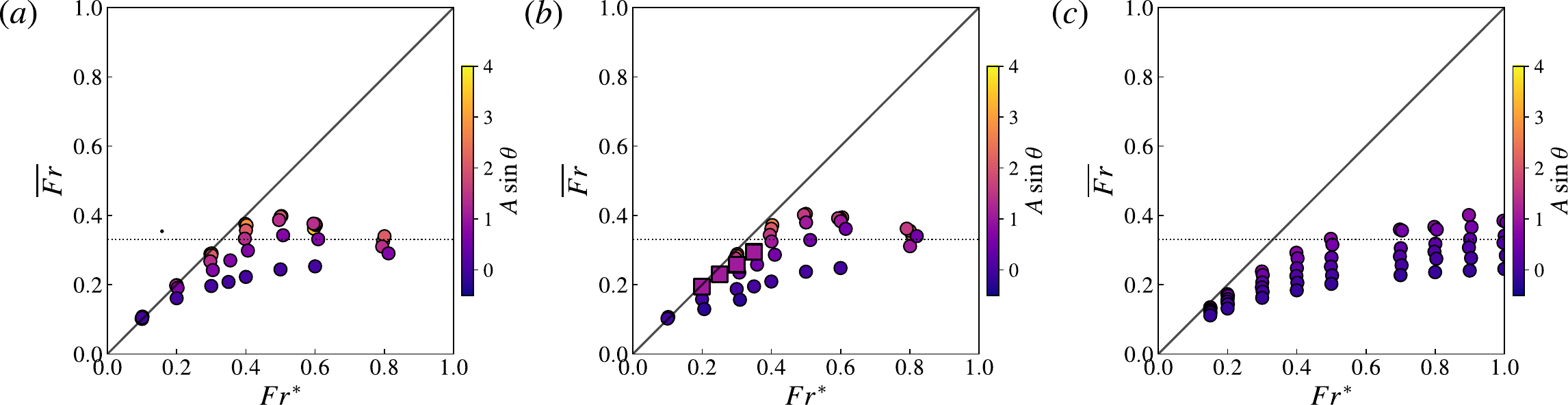}
    \caption{Comparison between the time-averaged Froude number $\overline{\Fr}$ from the numerical simulations and the theoretical laminar prediction $\Fr^*$ for (\textit{a}) wSID ($B=2$), (\textit{b}) mSID ($B=1$), and (\textit{c}) tSID ($B=1/4$). The solid black line marks $\overline{\Fr}=\Fr^*$, corresponding to the laminar HGV-A prediction. The dotted line marks the approximate wave-onset threshold $\overline{\Fr}=0.33$. Deviations below the solid line indicate that unsteady waves and mixing reduce the exchange flow relative to the laminar prediction.}
    \label{fig:FrversusFrstart}
\end{figure}

The analysis in figure~\ref{fig:fluctuations_changeFr} focuses on the mSID setup with $A\sin\theta=1.05$. Now, to examine wave onset across the entire simulation dataset, we compare the time-averaged Froude number $\overline{\Fr}$ from the numerical simulations with the laminar semi-analytical prediction $\Fr^*$. Figure~\ref{fig:FrversusFrstart} shows this comparison for (\textit{a}) wSID, (\textit{b}) mSID, and (\textit{c}) tSID. The time average is computed over $100\leq\check{t}\leq300$. The solid line marks the laminar prediction $\overline{\Fr}=\Fr^*$, while the dotted line marks the central LSA threshold $\overline{\Fr}=0.33$.

For small $\Fr^*$, the simulations follow the laminar prediction, which confirms the validity of the HGV-A approximation in the stable laminar regime. For larger $\Fr^*$, $\overline{\Fr}$ falls below $\Fr^*$, indicating that waves reduce the exchange flow relative to the laminar base state. Values above $\Fr^*\approx0.66$ should be interpreted cautiously because the width-averaged model does not resolve the fully 3D turbulent motions expected in real SID flows; in particular, the hydraulic limit $\overline{\Fr}=0.5$ is not recovered.

The four marked mSID cases in figure~\ref{fig:FrversusFrstart}(\textit{b}) correspond to the simulations shown in figure~\ref{fig:fluctuations_changeFr}. For $\Fr^*=0.20$, where no waves are present, the numerical flux satisfies $\overline{\Fr}\simeq\Fr^*$. For $\Fr^*=0.25$, waves are confined near the duct ends, but $\overline{\Fr}$ already falls below the laminar prediction. This shows that even localised Holmboe waves can reduce the net exchange flow. As $\Fr^*$ increases further, the waves penetrate further towards the duct centre and the deviation from $\overline{\Fr}=\Fr^*$ becomes larger.

The complete dataset shows the same trend predicted by the LSA. As $\Fr^*$ increases, $\overline{\Fr}$ departs from the laminar prediction, marking the dynamical effect of waves on the exchange flow. This departure occurs at larger $\Fr^*$ values when $A\sin\theta$ or $B$ is increased. As discussed in \S\ref{sec:interface}, both changes make the interior interface flatter and the local base flow more stable away from the duct ends. Thus, the flux reduction in figure~\ref{fig:FrversusFrstart} provides an integral signature of the same wave-onset mechanism identified by the LSA.

\section{Conclusions} \label{sec:conclusion}

We have derived a predictive framework for the onset of Holmboe waves in stratified exchange flows through long, inclined, laterally confined ducts at large Schmidt number. The central goal was to connect the imposed forcing and duct geometry to the laminar base flow and then to the threshold for interfacial-wave growth. This link has been missing from previous stratified inclined duct (SID) studies, where regime transitions were mapped experimentally and interpreted using measured profiles \citep{Lefauve2018}, prescribed profiles \citep{Ducimetiere2021EffectsInstabilities}, or 2D laminar theories \citep{Duran-Matute2023} that did not explicitly account for sidewall confinement.

The first component of the framework is a long-duct, sharp-interface asymptotic theory for the laminar exchange flow. This theory extends the 2D HGV-A (hydrostatic, gravitational, viscous-in-momentum, and advective-in-density) solution of \citet{Duran-Matute2023} to a 3D semi-analytical base flow that includes the effect of sidewalls. The solution predicts both the interface shape and the velocity field as functions of the control parameters $A\sin\theta$, $B$, and $\Rey_g$. It therefore provides a physically grounded base state for stability analysis, rather than relying on prescribed or measured profiles.

The second component is a width-averaged model in which the leading effect of sidewall friction is represented by an effective drag coefficient $C_D$, obtained from the 3D semi-analytical solution rather than empirically determined or assumed. This coefficient depends primarily on the spanwise aspect ratio $B$, with a weaker dependence on $A\sin\theta$. This parameterisation makes it possible to perform both linear stability analysis and high-$Sc$, time-dependent simulations over a broad parameter space. The simulations reproduce the laminar and Holmboe-wave regimes observed in prior laboratory SID experiments, while also capturing the main transition trends at a computational cost far below that of fully 3D simulations at $Sc\simeq700$, which remain prohibitive for scanning the parameter space.

The main result is that, for moderate and wide ducts, the laminar-to-Holmboe-wave transition collapses at a critical value $\Fr^* \simeq 0.33$, where $\Fr^*$ is the confinement-adjusted laminar Froude number computed from the semi-analytical base flow. This threshold is predicted by the LSA, observed in the width-averaged simulations, and is consistent with experimental regime diagrams over a wide range of $A\sin\theta, \Rey_g$, and $B$. Thus, rather than depending independently on forcing and confinement, wave onset is governed primarily by the laminar exchange state established by both, as measured by $\Fr^*$. 

The analysis also provides physical insight by showing that lateral confinement affects the onset problem in two distinct ways. First, it reshapes the laminar exchange flow, thereby changing the confinement-adjusted Froude number $\Fr^*$. Second, it directly damps unstable perturbations through sidewall friction, represented in the reduced stability problem by $C_D$. For moderate and wide ducts, the first mechanism dominates, so that wave onset is governed by an approximately constant critical value of $\Fr^*$. However, for sufficiently narrow ducts, direct damping also becomes important, causing the critical value of $\Fr^*$ to increase. Thus, the success and limitations of the proposed scaling have the same physical origin: $\Fr^*$ captures the influence of confinement on the laminar hydraulic state, but cannot account for the additional stabilising effect of sidewall friction acting directly on the instability.

The local stability analysis further shows that wave onset is not necessarily uniform along the duct. Although the central threshold is well described by $\Fr^*\simeq0.33$, waves can appear closer to the duct ends at lower values of $\Fr^*$, where the local base flow differs from that at the centre. The simulations show that such localised waves can reduce the net exchange flow even when the duct centre remains linearly stable. This result provides a physical explanation for the small differences between LSA thresholds, numerical classifications, and experimental regime boundaries, and highlights the importance of spatial variations in the local instability threshold.

The present model is designed primarily for the laminar-to-Holmboe-wave transition. It captures the onset of interfacial waves, but it is not expected to quantitatively describe the subsequent transition to intermittent turbulence, where secondary instabilities and highly nonlinear 3D motions become important \citep{Jiang2022}. In simulations and experiments, the next transition (Holmboe-wave to intermittent-turbulence) occurs near $\Fr^*\simeq0.66$ for $B=1$ and $B=2$ and at slightly higher values for $B=1/4$. The approximate factor of two between the laminar-to-wave and wave-to-turbulence thresholds is intriguing and remains an open question.

Overall, the results demonstrate that lateral confinement is not merely a secondary correction in stratified exchange flows. Instead, it fundamentally modifies both the laminar hydraulic state and the stability of that state. By combining asymptotic theory, sidewall-aware width averaging, linear stability analysis, high-Schmidt number simulations, and comparison with experiments, the framework provides the first coherent explanation for the previously observed dependence on $B$ of SID transition thresholds \citep{Lefauve2020,Duran-Matute2023}.

Although derived for the idealised laboratory SID flow, the underlying mechanisms are relevant to confined stratified exchange flows in estuaries, fjords, straits, man-made ducts and engineered systems, where sidewalls or lateral boundaries influence the onset of interfacial mixing.

\appendix
\section{Derivation of the semi-analytical base-flow solution}
\label{sec:AppendixSolution}
This appendix gives the algebraic details leading to the base-flow solution stated in \S\ref{sec:3Dsolution}. We first present the full dimensionless steady equations, because the long-duct reduction used in the main text follows directly from the powers of $A^{-1}$ appearing in these equations.

\subsection{Dimensionless equations and leading-order balance}
\label{app:dimensionless-equations}
Applying the scalings in \eqref{eq:scalings} to \eqref{eq:NS-before} gives the steady dimensionless equations
\begin{align}
A^{-1} \Rey_g \Fr\left(
    \tilde{u}\frac{\partial \tilde{u}}{\partial \tilde{x}}
    + \tilde{v}\frac{\partial \tilde{u}}{\partial \tilde{y}}
    + \tilde{w}\frac{\partial \tilde{u}}{\partial \tilde{z}}
\right)
&= -\frac{\partial \tilde{p}}{\partial \tilde{x}}
+ A^{-2}\frac{\partial^2\tilde{u}}{\partial \tilde{x}^2}
+ {B}^{-2} \frac{\partial^2 \tilde{u}}{\partial \tilde{y}^2}
+ \frac{\partial^2 \tilde{u}}{\partial \tilde{z}^2} \nonumber \\
&\quad + \frac{1}{4} \frac{\Rey_g}{ \Fr}\tilde{\rho}'\sin{\theta},
\label{eq:NS_u_AA} \\[1.0ex]
B^{2}A^{-3}\Rey_g \Fr\left(
    \tilde{u}\frac{\partial \tilde{v}}{\partial \tilde{x}}
    + \tilde{v}\frac{\partial \tilde{v}}{\partial \tilde{y}}
    + \tilde{w}\frac{\partial \tilde{v}}{\partial \tilde{z}}
\right)
&= -\frac{\partial \tilde{p}}{\partial \tilde{y}}
+ A^{-4}B^{2}\frac{\partial^2\tilde{v}}{\partial \tilde{x}^2}
+ A^{-2}\frac{\partial^2 \tilde{v}}{\partial \tilde{y}^2}
+ A^{-2}B^2\frac{\partial^2 \tilde{v}}{\partial \tilde{z}^2},
\label{eq:NS_v_AA} \\[1.0ex]
A^{-3}\Rey_g \Fr\left(
    \tilde{u}\frac{\partial \tilde{w}}{\partial \tilde{x}}
    + \tilde{v}\frac{\partial \tilde{w}}{\partial \tilde{y}}
    + \tilde{w}\frac{\partial \tilde{w}}{\partial \tilde{z}}
\right)
&= -\frac{\partial \tilde{p}}{\partial \tilde{z}}
+ A^{-4}\frac{\partial^2\tilde{w}}{\partial \tilde{x}^2}
+ A^{-2}B^{-2}\frac{\partial^2 \tilde{w}}{\partial \tilde{y}^2}
+ A^{-2}\frac{\partial^2 \tilde{w}}{\partial \tilde{z}^2} \nonumber \\
&\quad -\frac{1}{4} \frac{\Rey_g}{\Fr}A^{-1}\tilde{\rho}'\cos{\theta},
\label{eq:NS_w_AA} \\[1.0ex]
\frac{\partial \tilde{u}}{\partial \tilde{x}}
+ \frac{\partial \tilde{v}}{\partial \tilde{y}}
+ \frac{\partial \tilde{w}}{\partial \tilde{z}}
&= 0,
\label{eq:NS_cont_AA} \\[1.0ex]
A^{-1}\Rey_g Sc \Fr\left(
    \tilde{u}\frac{\partial \tilde{\rho}'}{\partial \tilde{x}}
    + \tilde{v}\frac{\partial \tilde{\rho}'}{\partial \tilde{y}}
    + \tilde{w}\frac{\partial \tilde{\rho}'}{\partial \tilde{z}}
\right)
&= A^{-2}\frac{\partial^2 \tilde{\rho}'}{\partial \tilde{x}^2}
+ B^{-2}\frac{\partial^2 \tilde{\rho}'}{\partial \tilde{y}^2}
+ \frac{\partial^2 \tilde{\rho}'}{\partial \tilde{z}^2}.
\label{eq:AD_AA} 
\end{align}
Taking the limit $A\to\infty$ and retaining the leading-order terms gives \eqref{eq:hydrostatic_x}--\eqref{eq:Transport_assym} in the main text. 

\subsection{Solution of the forced Poisson problem}
\label{app:poisson-solution}

We now solve \eqref{eq:hydrostatic_x}--\eqref{eq:Transport_assym} for the two-layer density field \eqref{eq:density}. In each layer, indexed by $\zeta=\pm1$, integration of the hydrostatic balance \eqref{eq:Hydrostatic_z} and continuity of pressure across $\tilde{z}=\eta(\tilde{x})$ give
\begin{equation}
    \tilde{p}_{0,\zeta}(\tilde{x}, \tilde{z})=\zeta\frac{K}{4}[\cos{\theta}(\tilde{z}-\eta(\tilde{x})-\zeta)+\zeta f(\tilde{x})],
\end{equation}
with $f(\tilde{x})$ still unknown \citep[more detailed steps are given by][]{Duran-Matute2023}. 
Substituting this pressure field into \eqref{eq:hydrostatic_x} yields
\begin{equation}
    \frac{\partial^2 \tilde{u}_{0,\zeta}}{\partial \tilde{z}^2}+B^{-2}\frac{\partial^2 \tilde{u}_{0,\zeta}}{\partial \tilde{y}^2} = -\zeta \frac{K}{4} F_\zeta (\tilde{x}),
    \label{eq:poisson}
\end{equation}
where  
\begin{equation}
    F_\zeta(\tilde{x})=\cos{\theta} \frac{\mathrm{d}\eta}{\mathrm{d}\tilde{x}}-\zeta\frac{\mathrm{d}f}{\mathrm{d}\tilde{x}}-A\sin{\theta}.
\end{equation}
The remaining task is therefore to solve this forced Poisson problem for the streamwise velocity in each layer with the forcing $K F_\zeta(\tilde{x})/4$.

The boundary conditions are
\begin{equation}
    \tilde{u}_{0,\zeta}=0
    \quad\text{at}\quad
    \tilde{y}=\pm1,\quad
    \tilde{z}=\eta(\tilde{x}),\quad
    \text{and}\quad
    \tilde{z}=\zeta,
\end{equation}
representing no-slip boundary conditions at the duct walls, and no velocity at the interface. 

Equation~\eqref{eq:poisson} is solved separately in each layer by decomposing the solution into a particular solution $\chi$ and two homogeneous corrections $\phi_1$ and $\phi_2$,
\[
    \tilde{u}_{0,\zeta}=\chi-\phi_1-\phi_2.
\]
The particular solution $\chi$ satisfies the forced equation, while $\phi_1$ and $\phi_2$ enforce the no-slip conditions on the two sidewalls. For the lower layer, the particular solution is
\begin{equation}
    \chi=\frac{K}{8}F_\zeta(\tilde{x})(\tilde{z}+1)(\tilde{z}-\eta(\tilde{x})),
\end{equation}
and
\begin{equation}
    \phi_1 = \sum_{n=1}^\infty a_n \sin{\left[\frac{n\pi (\tilde{z}+1)}{1+\eta}\right]}\sinh{\left[\frac{nB\pi(1+\tilde{y})}{1+\eta}\right]},
\end{equation}
where $a_n$ can be found applying the boundary conditions at $\phi_1(\tilde{y}=+1)$ and taking $b_n=a_n \sinh{\left(\frac{2nB\pi}{1+\eta}\right)}$, which yields 


\begin{equation}
    b_n=\frac{K F_\zeta(\tilde{x})}{4(\eta+1)} \int^\eta_{-1}(\tilde{z} + 1)(\tilde{z} - \eta(\tilde{x}))\sin{\left[\frac{n\pi(\tilde{z}+1)}{\eta+1}\right]}\mathrm{d}\tilde{z}
    =\frac{K F_\zeta(\tilde{x})}{2\pi^3n^3}[(-1)^n-1](\eta+1)^2.
\end{equation}

It can easily be seen that this is only non-zero for odd values of $n$. Furthermore, the solution for $\phi_2$ is just equal to $\phi_1$ mirrored about $\tilde{y}=0$. Combining $\chi$, $\phi_1$, and $\phi_2$ yields the solution for $\tilde{u}_0$ given in \eqref{eq:EquationForuwithn}.

To fully specify $\tilde{u}_0$, we still need to obtain $F_\zeta(\tilde{x})$, which in turn requires evaluating $\mathrm{d}f/\mathrm{d}\tilde{x}$ and $\mathrm{d}\eta/\mathrm{d}\tilde{x}$. We first apply the condition of zero mean flow through the duct:
\begin{equation}
    \int_{-1}^\eta\int^1_{-1} \tilde{u}_{0,-1}(\tilde{x},\tilde{y},\tilde{z})\,\mathrm{d}\tilde{y}\,\mathrm{d}\tilde{z} +\int_\eta^1\int^1_{-1}\tilde{u}_{0,+1}(\tilde{x},\tilde{y},\tilde{z})\,\mathrm{d}\tilde{y}\,\mathrm{d}\tilde{z}=0,
    \label{eq:zeromean}
\end{equation}
yielding an equation for $\mathrm{d}f/\mathrm{d}\tilde{x}$:
\begin{equation}
    \frac{\mathrm{d}f}{\mathrm{d}\tilde{x}}=\frac{P_{+1}(\tilde{x})(-1+\eta)^3-P_{-1}(\tilde{x})(1+\eta)^3}{P_{-1}(\tilde{x})(1+\eta)^3+P_{+1}(\tilde{x})(1-\eta)^3}\left(\frac{\mathrm{d}\eta}{\mathrm{d}\tilde{x}}\cos{\theta}-A\sin{\theta}\right),
    \label{eq:dfdx}
\end{equation}
with
\begin{equation}
    P_\zeta(\tilde{x}) = -\frac{\pi^4}{96}+\frac{1-\zeta\eta}{B\pi}\sum^\infty_{n=1}\frac{1}{(2n-1)^5}\tanh{\left[\frac{B(2n-1)\pi}{1-\zeta\eta}\right]}.
    \label{eq:Hzeta}
\end{equation}
In the main text, we retain only the dominant $n=1$ contribution to $P_\zeta$, since the higher-order terms are suppressed by the factor $(2n-1)^{-5}$. The remaining unknown, $\mathrm{d}\eta/\mathrm{d}\tilde{x}$, is obtained by imposing the prescribed dimensionless flux in each layer,
\begin{equation}
    \int_{-1}^\eta\int^1_{-1} \tilde{u}_{0,-1}(\tilde{x},\tilde{y},\tilde{z})\mathrm{d}\tilde{y}\mathrm{d}\tilde{z} = 2 = -\int_\eta^1\int^1_{-1}\tilde{u}_{0,+1}(\tilde{x},\tilde{y},\tilde{z})\mathrm{d}\tilde{y}\mathrm{d}\tilde{z},
    \label{eq:Finddetadx}
\end{equation}
which gives \eqref{eq:detadx}. Substitution into \eqref{eq:dfdx} then gives the forcing amplitude $F_\zeta$ used in \eqref{eq:Fx}.

\section{Validation of the width-averaged simulations against the semi-analytical laminar solution} \label{sec:AppendixAnalyticalComp}

\begin{figure}
    \includegraphics[width=0.9\textwidth]{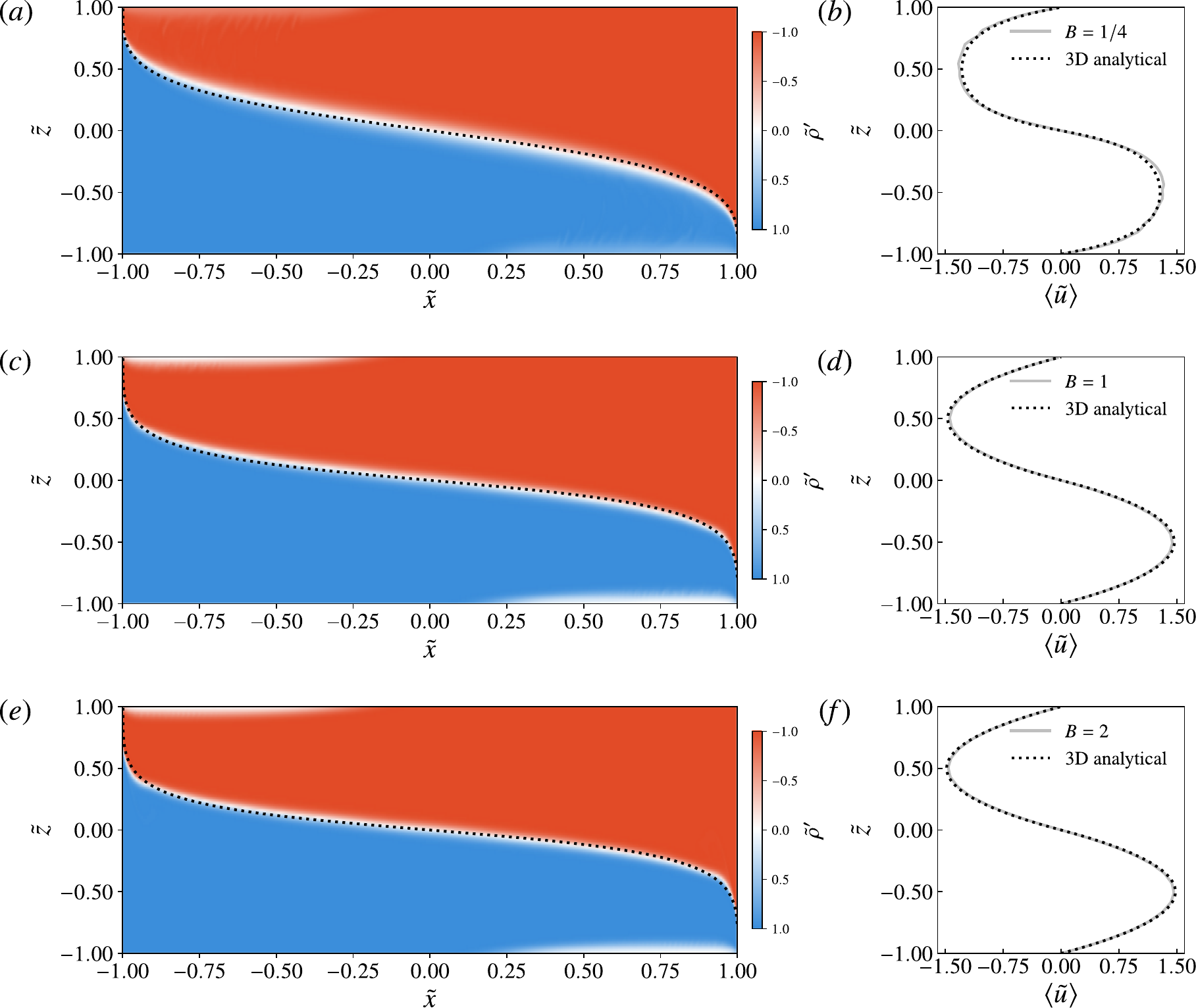}
    \caption{Comparison of the flow characteristics between the width-averaged numerical simulation and the 3D semi-analytical solution for laminar flow with $(\Rey_g,\ A\sin \theta) = (200,  1.05)$ and different spanwise aspect ratios $B$. (\textit{a}), (\textit{c}), and (\textit{e}) show the density fields for $B = 1/4$, $1$, and $2$, respectively, together with the interface location $\eta(\tilde{x})$ obtained from the semi-analytical solution, shown as dotted black curves. (\textit{b}), (\textit{d}), and (\textit{f}) present the corresponding width-averaged velocity profiles at the duct centre $(\tilde{x} = 0)$. The numerical results are shown as solid curves and analytical predictions as dotted curves.}
    \label{fig:LaminarComp}
\end{figure}

Figure~\ref{fig:LaminarComp} compares the width-averaged simulations with the semi-analytical laminar solution for three spanwise aspect ratios, $B=1/4$, $1$, and $2$. The remaining parameters are chosen to match the experimental configuration of \citet{Lefauve2019}: $\Rey_g=200$, $\theta=2^\circ$, $A=30$, and $Sc=700$. Figures~\ref{fig:LaminarComp}(\textit{a}), (\textit{c}), and (\textit{e}) show the width-averaged density fields from the numerical simulations. The light region marks the interface, where $\tilde{\rho}'=0$, and the dotted black curve shows the analytical prediction $\eta(\tilde{x})$. The agreement is good in all cases, both in the interface position and in its overall slope. As the duct width increases, the interface becomes flatter near the duct centre, consistent with the behaviour described in \S\ref{sec:interface}.

To extend the comparison, we present the velocity profile at the duct centre $(\tilde{x} = 0)$ in figures~\ref{fig:LaminarComp}(\textit{b}), (\textit{d}), and (\textit{f}) for $B = [1/4,\ 1,\ 2]$, respectively. For the analytical solution, the results shown as dotted curves represent the width-averaged values obtained using the 3D analytical solution. The width-averaged simulation is shown as a solid line. The influence of lateral confinement is evident in these results: smaller spanwise aspect ratios lead to a deceleration of the flow and a flattening of the velocity profile. This is attributed to the higher wall shear stress induced by confinement. As $B$ increases, a corresponding increase in maximum velocities is observed, close to $\tilde{z} = \pm 0.5$. Note that the shear stress at the interface increases with larger values of $B$, further steepening the interface. Generally, the agreement is excellent between the numerical simulation and the analytical solution for the base flow.

\section{Comparison between width-averaged simulations and laboratory experiments}\label{sec:AppendixExperimentalComp} 

\begin{figure}
    \includegraphics[width=\textwidth]{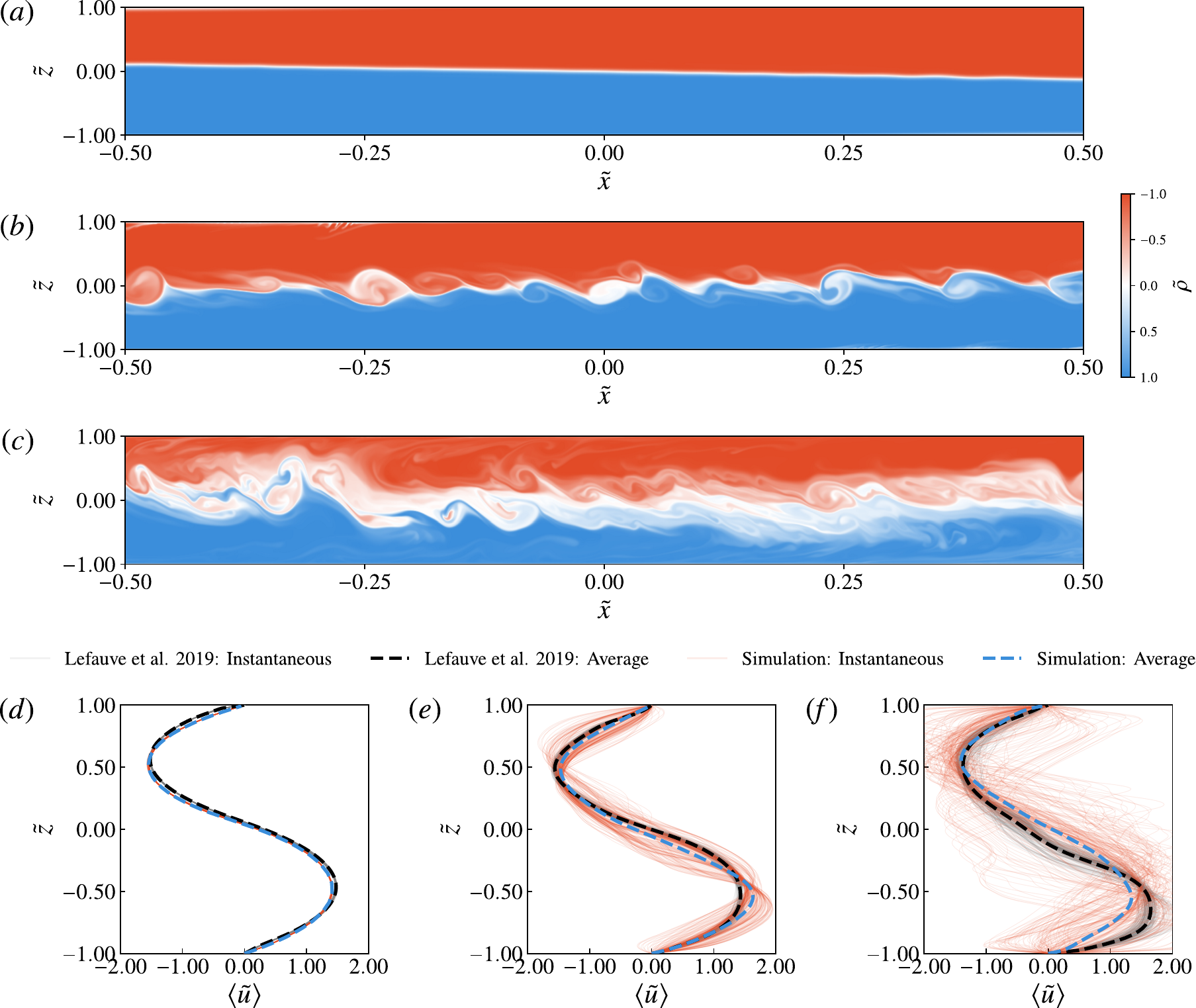}
    \caption{Comparison between width-averaged simulations and laboratory observations for representative mSID regimes. Panels (\textit{a})--(\textit{c}) show simulated density fields for laminar flow $(\Rey_g,A\sin\theta)=(360,1.05)$, Holmboe waves $(990,1.05)$, and intermittent turbulence $(2024,1.05)$. Panels (\textit{d})--(\textit{f}) compare width-averaged velocity profiles at $\tilde{x}=-0.3$ with experimental measurements for the corresponding regimes, using experimental parameters $(398,1.05)$, $(1059,1.05)$, and $(2024,1.05)$. Dashed curves denote time averages and lighter curves denote instantaneous profiles.}
    \label{fig:ExperimentalComp}
\end{figure}

Figure~\ref{fig:ExperimentalComp} compares representative width-averaged simulations with laboratory observations for the mSID configuration, $A=30$ and $B=1$, using the experimental dataset of \citet{Lefauve2019b}. The cases are chosen to represent the laminar, Holmboe-wave, and intermittently turbulent regimes. The corresponding experimental points are $(\Rey_g,A\sin\theta)=(398,1.05)$, $(1059,1.05)$, and $(2024,1.05)$. Because the numerical transition thresholds differ slightly from the experimental ones, the laminar and Holmboe-wave simulations are shown at $(360,1.05)$ and $(990,1.05)$, respectively. The intermittently turbulent case is shown at the same parameter value as the experiment, $(2024,1.05)$, since it lies further away from the transition boundary.

Figures~\ref{fig:ExperimentalComp}(\textit{a})--(\textit{c}) show snapshots of the simulated density field for the three regimes. The laminar and Holmboe-wave regimes are clearly reproduced. The intermittently turbulent case displays disordered structures over a range of length scales and enhanced mixing across the duct height. It also reveals the expected limitation of 2D simulations: coherent vortical structures are more persistent than in experiments because vortex stretching is absent in 2D flows \citep[e.g.][]{Boffetta2012}.

Figures~\ref{fig:ExperimentalComp}(\textit{d})--(\textit{f}) compare width-averaged velocity profiles at $\tilde{x}=-0.3$ between experiments and simulations. Dashed curves show time-averaged profiles, computed for $100\leq\check{t}\leq300$ for the simulations and using all available times for the experiments. Lighter curves show instantaneous profiles, sampled every $5\check{t}$ in the simulations and at all available experimental times. For the laminar and Holmboe-wave cases, the mean profiles agree well. Some differences appear in the instantaneous profiles, especially in the wave and intermittently turbulent regimes, where the 2D simulations produce larger and longer-lived vortical structures than those observed experimentally. In the intermittently turbulent case, the mean profiles remain qualitatively similar, but the instantaneous fluctuations are stronger in the simulations. We also note that the experimental profiles do not exactly satisfy the zero-mean-flow condition in \eqref{eq:zeromean}, which partially explains the differences between the experiments and the simulations.

\backsection[Acknowledgements]{G.S. de Aquino thanks Antoon van Hooft for his assistance with the initial Basilisk implementation and Zhonghan Xue for his contributions to code improvements.}

\backsection[Funding]{G.S. de Aquino's research stay at Eindhoven University of Technology was co-funded by the Erasmus+ programme of the European Union and the ED MEGeP mobility grant through the Université de Toulouse. A. Lefauve was supported by a NERC Independent Research Fellowship NE/W008971/1. The computational resources were provided through an NWO Domain Science grant (2024.006), and the simulations were carried out at the Dutch National Supercomputing Facility SURF.}

\backsection[Declaration of Interests]{The authors declare none.}


\bibliographystyle{jfm}
\bibliography{references_clean}
\end{document}